\documentstyle[aps,prb,multicol,epsf]{revtex}

\newcommand{\equ}[1]{Eq.~(\protect\ref{#1})}
\newcommand{\equs}[2]{Eqs.~(\protect\ref{#1})-(\protect\ref{#2})}
\newcommand{\bfm}[1]{\bbox{#1}}

\renewcommand{\caption}[1]{\refstepcounter{figure}\protect\noindent%
  \protect\parbox{8.6cm}{\small FIG. \thefigure. #1}}

\begin{document}

\draft

\title{Elasticity and melting of vortex crystals in anisotropic
superconductors:\\ 
Beyond the continuum regime}

\author{M.-Carmen Miguel$^{1,2}$ and Mehran Kardar$^{1}$}

\address{$^1$Physics Department, Massachusetts Institute of Technology\\ 
Cambridge, Massachusetts 02139, USA\\
$^2$The Abdus Salam ICTP\\
P.O. Box 586, 34100 Trieste, Italy}

\date{\today}

\maketitle

\begin{abstract}

The elastic moduli of vortex crystals in anisotropic superconductors
are frequently involved in the investigation of their phase diagram
and transport properties.  We provide a detailed analysis of the
harmonic eigenvalues (normal modes) of the vortex lattice for general
values of the magnetic field strength, going beyond the elastic
continuum regime.  The detailed behavior of these wavevector-dependent
eigenvalues within the Brillouin zone (BZ), is compared with several
frequently used approximations that we also recalculate.  Throughout
the BZ, transverse modes are less costly than their longitudinal
counterparts, and there is an angular dependence which becomes more
marked close to the zone boundary.  Based on these results, we propose
an analytic correction to the nonlocal continuum formulas which fits
quite well the numerical behavior of the eigenvalues in the London
regime.  We use this approximate expression to calculate thermal
fluctuations and the full melting line (according to Lindeman's
criterion) for various values of the anisotropy parameter.
 
\end{abstract}

\begin{multicols}{2}

\section{Introduction}
\label{sec:intro}

\begin{figure}
\epsfxsize=8truecm{\epsfbox{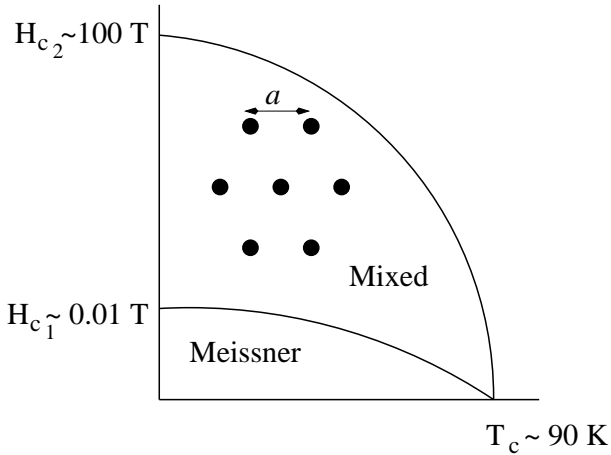}}
\medskip
\narrowtext{\caption{Schematic mean-field phase diagram of a type-II
superconductor. The numbers are indicative orders of magnitude for
highly anisotropic materials like ${\rm YBa_2Cu_3O_7}$ and ${\rm
Bi_2Sr_2CaCu_2O_8}$.\label{mfphases} }}
\end{figure}

Among the fascinating aspects of the new high-temperature
superconducting materials, there are many which are due to the rich
and complex behavior of vortex lines \cite{Blatter94}.  These quanta
of magnetic flux penetrate the superconductor above a certain
threshold value of the external magnetic field $H$, the so-called {\em
lower critical field} $H_{c_1}$ (roughly $10^{-2} T$ , and their
concentration increases with $H$ up to the {\em upper critical field}
$H_{c_2}$ (approximately $10^2 T$), above which the material is
normal\cite{Tinkham,numbers} (see Figure~\ref{mfphases}).  In a
classic work, Abrikosov showed that the minimum energy arrangement of
flux lines in a conventional superconductor is a triangular lattice,
with a lattice spacing $a$ which varies with the magnetic field
strength\cite{Abrikosov57}.

Recently, with the discovery of high temperature cuprate
superconductors, there has been a resurgence of interest in the
properties of vortex matter.  One novel aspect of these materials is
the symmetry of the superconducting order parameter. A series of
experiments \cite{Harlingen95} seem to provide evidence consistent
with $d$-wave symmetry, in contrast to conventional superconductors
which have $s$-wave symmetry.  In addition to modifying the internal
structure of the vortex lines, $d$-wave symmetry has implications for
the global mean-field arrangement of the vortex lattice \cite{BWXF}.
In particular, an oblique square lattice appears to replace the
standard triangular arrangement, as the most stable configuration in
part of the $B-T$ phase diagram \cite{KeMa}.  (The new phase appears
for high values of the magnetic field, whereas the triangular
arrangement is confined to low fields.)

Furthermore, the relative flexibility of vortex lines in these
materials makes them susceptible to distortions by thermal
fluctuations, and other sources of disorder (oxygen impurities, grain
boundaries, etc.).  The regular lattices obtained in mean-field theory
are thus distorted, giving rise to a rich variety of vortex-matter
phases\cite{Nelson88,Fisher91}.  It is thus necessary to understand
the elastic response of vortex lattices to distortions, a subject that
has been intensely studied in the context of conventional
superconductors.  The stability of the triangular lattice against
small distortions is guaranteed as long as the characteristic length
of the field variations, the so-called {\em penetration depth}
$\lambda$, remains smaller than the size of the system
\cite{Fetter66}.  (An infinite $\lambda$ renders the lattice of vortex
lines unstable against fluctuations or, more precisely, against shear
deformations.)  At low temperatures, fluctuations are well described
by small corrections to the mean-field results. Although the
phenomenology is similar for $d$-wave superconductors, less is known
about the effects of thermal fluctuations and disorder in this case.
We shall thus focus on a triangular lattice of vortex lines. The
extension of the results to oblique configurations is possible, and
should prove interesting.

The elastic properties of the triangular vortex lattice at long
distances are characterized by its compressional, shear, and tilt
moduli.  These moduli are frequently involved in the theoretical and
experimental determination of the properties of the material, as for
example, the intricate vortex phase diagrams.  As such, the derivation
of the elastic energy and the elastic moduli has been undertaken in
several publications
\cite{Matricon64,Brandt77,Nelson89,Kogan89,Brandt89,Houghton89,Sudbo91a,Sudbo91b,Fisher-rev91,Sardella92}.
While it is well known that the elastic moduli of the vortex lattice
depend strongly on the magnetic field strength, the values currently
presented in the literature are only strictly useful in certain
limiting situations.  Most frequently, results are obtained in the
so-called {\em continuum limit}, in which one disregards the discrete
nature of the underlying vortex lattice.  Obviously, this description
is not suitable for the whole range of possible flux-line densities in
the mixed state, which extends from $H_{c_1}$ up to the upper critical
field $H_{c_2}$, at which the magnetic field penetrates uniformly into
the material.  The elastic properties of the vortex lattice, and hence
its stability against thermal fluctuations, depend sensitively on the
value of the magnetic field in this range.  Furthermore, the important
fluctuations sometimes occur at short wavelengths, where a simple
elastic description may not be appropriate.  It is thus worthwhile to
obtain the general dependence of the energy cost of harmonic
distortions of the vortex lattice.

As temperature is increased, thermal fluctuations cause the vortex
lattice to {\em melt} into a vortex liquid.  Several experiments
employing quite different techniques have provided firm evidence for
such a transition
\cite{Gammel88,Safar92,Kwok92,Pastoriza95,Zeldov95,Liang96,Welp96,Roulin96,Oral98}.
In the new cuprate superconductors, the melting transition can occur
at temperatures well below the mean field point, so that the Abrikosov
lattice is melted over a substantial portion of the phase diagram.
Furthermore, the vortex lattice can melt not only by increasing
temperature, but also by decreasing the magnetic field to the vicinity
of $H_{c_1}(T)$. In this region of the phase diagram, the
concentration of vortices is very dilute; the separation $a$ between
neighboring flux lines is larger than the penetration length
$\lambda$, and the vortex-vortex interaction decays exponentially.  As
a consequence, the elastic moduli become exponentially small, and
correspondingly, thermal fluctuations are greatly enhanced. This
behavior gives rise to an interesting reentrant behavior of the
melting line at low fields.
  
In this paper, we take up the computation of the elastic properties of
a vortex lattice, in a systematic manner which is not restricted to
the most commonly considered continuum limit.  Our analysis allows the
illustration of some of the subtleties involved in the calculation,
and to successfully resolve some of the unclear points that we
encountered in our review of the literature, and which may also have
puzzled other investigators of the subject.  The paper is organized as
follows: In Sec.~\ref{sec:general}, we introduce the Ginzburg-Landau
Hamiltonian for an anisotropic superconductor.  This model is commonly
used in the literature as the starting point for studying the effects
of fluctuations and disorder.  Minimization of this Hamiltonian gives
the optimal lattice configuration, around which we then study the cost
of distortions in the harmonic approximation.  This section also
introduces the main parameters and notation used throughout the paper.
In Sec.~\ref{sec:moduli}, we provide an expression for the harmonic
kernel in terms of a sum over Bravais lattice vectors or,
equivalently, over reciprocal lattice vectors. The normal mode
eigenvalues are explicitly written in terms of this
kernel. Sec.~\ref{sec:limits} is composed of subsections in which we
introduce certain limiting situations, corresponding to a single line,
local, local--continuum, and nonlocal--continuum limits.  These
limiting cases are often quoted in the literature, and are widely used
in studies of vortex matter in high temperature superconductors.
Within our framework, we recalculate the analytic values of the
elastic eigenvalues in these limits before discussing our more general
results in Section \ref{sec:numeric}, where the harmonic eigenvalues
are numerically evaluated.  Our results are graphically presented in
several plots, and compared to the limits introduced in
Sec.~\ref{sec:limits}, to easily visualize the accuracy of the
approximations involved.  As a practical illustration of the potential
applications of our results, we use the harmonic eigenvalues to
calculate the thermal distortions of the vortex lattice in Section
\ref{sec:melting}.  The leading contribution to flux-line fluctuations
in real space comes from the transverse modes.  In conjunction with
the Lindeman's criterion, we can then find the full form of the
melting line as a function of the magnetic field.  This is one of many
potential applications that our general analysis makes possible,
without the need to extrapolate the elastic behavior of the lattice to
regimes beyond their range of validity.  Finally, in the last section
we summarize our main conclusions.

\section{General formulation}

\label{sec:general}

Our starting point is the continuum Ginzburg-Landau free energy for an
anisotropic superconductor in the London limit \cite{Tinkham}.  In
this limit, the penetration length $\lambda$ ($\sim 10^3$ \AA) is much
larger than the coherence length of the superconductor $\xi$ ($\sim
10$ \AA), and fluctuations in the magnitude of the order parameter
$\Psi_o$ are neglected.  The phase degree of freedom is then the only
variable that needs to be considered.  For the anisotropic
superconductors under consideration, this approximation breaks down in
a narrow band close to $H_{c_2}$, where the separation between
vortices becomes comparable to $\xi$.  The Ginzburg-Landau Hamiltonian
in this limit is given by
\begin{eqnarray}
\label{eq:g1}
{\cal H}&=&\int d^{3} \bbox{r} \left[\alpha\left(\nabla_{\perp} 
\theta-\frac{2\pi}{\phi_o} \bbox{A}_{\perp}\right)^2
\right.\\ \nonumber 
&+& \left. \alpha\ 
\epsilon^2 \left(\partial_z \theta-\frac{2\pi}{\phi_o} A_z\right)^2 + 
\frac{\bbox{b}^2}{8\pi}-\frac{\bbox{b}\cdot\bbox{H}}{4\pi}\right].
\end{eqnarray}
Here $\theta(\bbox{r}_{\perp},z)$ is the phase of the complex order
parameter field $\Psi(\bbox{r}_{\perp},z)=\Psi_o
e^{i\theta(\bbox{r}_{\perp},z)}$, $\bbox{A}(\bbox{r}_{\perp},z)$ is
the magnetic vector potential related to the magnetic induction
through $\bbox{b}(\bbox{r}_{\perp},z)=
\nabla\times\bbox{A}(\bbox{r}_{\perp},z)$, $\alpha=(\hbar
\Psi_o)^2/2m$, $\phi_o=hc/2e$ is the flux quantum, and
$\epsilon^2=m/M$ is the usual anisotropy parameter, with $m$ and $M$
being the effective masses in the Cu-O planes of the material, and
along the perpendicular $c$ axis chosen to coincide with the
$z$ direction, respectively. The external magnetic field ${\bfm H}$ is
also oriented parallel to the $z$ axis ${\bfm H}=H\hat{e}_z$. Note
that $\nabla_{\perp}$, $\bbox{r}_{\perp}$, and $\bbox{A}_{\perp}$,
denote the planar components of $\nabla$, $\bbox{r}$, and $\bbox{A}$,
respectively.

Consider a configuration with $N$ vortex lines, and decompose the
corresponding phase field into two parts:
$\sum_{n=1}^{N}\theta^s_n[\bbox{r}_{\perp}-\bbox{R}_{n,\perp}(z),z]$,
which represents the singular phase due to the $N$ vortices passing
through points $\bbox{R}_n(z)= \bbox{R}_{n,\perp}(z)+z\hat{z}$
$(n=1,\ldots,N)$ on independent planes at each $z$; and $\theta^r$, a
regular phase field accounting for the relaxation due to the couplings
between the planes. By construction, each $\theta^s_n$ is the solution
of a two-dimensional problem with the circulation constraint
$\oint_{\cal C} d \theta^s_n=2\pi$ on any closed circuit ${\cal C}$
around the $n$-th vortex.  We have chosen the coordinate $z$ to
parametrize the trajectories of the different flux
lines. Nonetheless, one has to bear in mind that the results should be
invariant under an arbitrary reparametrization.

Variations of Eq. (\ref{eq:g1}) with respect to the phase $\theta^r$,
and the vector potential $\bbox{A}$, provide the differential
equations for these quantities, whose solutions minimize the energy in
(\ref{eq:g1}).  After considering the Coulomb gauge
$\nabla\cdot\bbox{A}=0$, these equations read
\begin{equation}\label{eq:g2}
(\Delta_{\perp}+\epsilon^2\partial_z^2)\theta^r-
\frac{2\pi}{\phi_o}(\epsilon^2-1)\partial_z A_z= 
-\epsilon^2\partial_z^2\sum_{n=1}^{N}\theta^s_n,
\end{equation}

\begin{equation}\label{eq:g4}
\lambda^2 \Delta A_z-\epsilon^2 A_z =- 
\epsilon^2\frac{\phi_o}{2\pi}\partial_z\left(\theta^r+ 
\sum_{n=1}^{N}\theta^s_n\right),
\end{equation}
\begin{equation}\label{eq:g3}
\lambda^2 \Delta \bbox{A}_{\perp}-\bbox{A}_{\perp}= 
-\frac{\phi_o}{2\pi}\nabla_{\perp}\left(\theta^r+   
\sum_{n=1}^{N}\theta^s_n\right),
\end{equation}
where we have introduced the squared penetration depth in a plane
perpendicular to the $z$ axis,
$\lambda^2=\phi_o^2/(32\pi^3\alpha)=mc^2/(16\pi e^2 \Psi_o^2)$, and
$\Delta_{\perp}$ stands for the in-plane component of the Laplacian
operator.  The coupled equations (\ref{eq:g2}) and (\ref{eq:g4}) are
easily solved in Fourier space to give
\begin{equation}\label{eq:g5}
\theta^r(\bbox{k}) = 
\frac{i k_z (\lambda^2 k^2+1)}{k^2 (\lambda_c^2 q^2+\lambda^2 k_z^2 +1)}
\sum_{n=1}^{N} {\cal F}_n(\bbox{k}),
\end{equation}
\begin{equation}\label{eq:g7}
A_z(\bbox{k}) =\frac{\phi_o}{2\pi}
\frac{q^2}{k^2 (\lambda_c^2 q^2+\lambda^2 k_z^2 +1)}
\sum_{n=1}^{N} {\cal F}_n(\bbox{k}),
\end{equation}
where $\bbox{k}=\bbox{q}+k_z\hat{z}$, $\lambda_c=\lambda/\epsilon$ is
the penetration length along the c-axis of the superconductor, and we
have introduced the function
\begin{equation}
{\cal F}_{n}(\bbox{k})=
-\frac{2\pi i}{q^2}(\bbox{q}\times\hat{z})\cdot 
\int  dz\ {\rm e}^{ik_z z}\ {\rm e}^{i\bbox{q}\cdot
\bbox{R}_{n,\perp}(z)} 
\frac{d\bbox{R}_{n,\perp}(z)}{dz},\label{eq:g9}
\end{equation}
which represent the Fourier transform of the vortex function
$\partial_z\theta_n^s$. 

Because of its singular nature, the Fourier transform of  
$\nabla_{\perp}\theta_n^s$, actually has a component
\begin{equation}
{\bfm P}_{n}(\bbox{k}) = \frac{2\pi
 i}{q^2}(\bbox{q} 
 \times\hat{z})
 \int  dz\ {\rm e}^{ik_z z}\ {\rm e}^{i\bbox{q}\cdot
 \bbox{R}_{n,\perp}(z)}, \label{eq:g10}
\end{equation}
which is perpendicular to ${\bf q}$. 
We thus define the transversal and longitudinal components of
the in-plane projection of the gauge field, as 
${\bfm   A}_{\perp}^T=(I-\hat{q}\hat{q})\cdot{\bfm A}_{\perp}$, and 
${\bfm   A}_{\perp}^L=\hat{q}\hat{q}\cdot{\bfm A}_{\perp}$ respectively.
The former is obtained from \equ{eq:g3} as
\begin{equation}\label{eq:g6}
\bbox{A}_{\perp}^{T}(\bbox{k}) = 
\frac{\phi_o}{2\pi(\lambda^2 k^2+1)} \sum_{n=1}^{N}\bbox{P}_n({\bfm k}),
\end{equation}
while the latter is obtained from the Coulomb gauge condition as
\begin{equation}\label{eq:g8}
\bbox{A}_{\perp}^{L}(\bbox{k}) = -\frac{k_z\bbox{q}}{q^2} A_z(\bbox{k}).
\end{equation}

For a given distortion of the flux-lattice, the above solutions provide
the form of the phase and gauge fields that minimize \equ{eq:g1}.
The next step is to evaluate the energy cost of such distortions
 by substituting these solutions into the Ginzburg-Landau free energy, 
resulting in
\begin{eqnarray}\label{eq:g11}
{\cal H}&=&\frac{\phi_o^2}{32\pi^3}\int \frac{
d^{3}\bbox{k}}{(2\pi)^3} \left\{\frac{q^2}{(\lambda_c^2 q^2
+\lambda^2 k_z^2 +1)} \left|\sum_{n=1}^N{\cal F}_n\right|^2
\right.\nonumber \\ &+& \left.\frac{k^2}{(\lambda^2 k^2
+1)}\left|\sum_{n=1}^N{\bfm P}_n\right|^2 \right\}-L\frac{\phi_o N
H}{4\pi}.
\end{eqnarray}
Let us now assume that the position vector $\bfm{R}_{n,\perp}(z)$ is
the sum of a perfect lattice vector, plus a small displacement field
due to fluctuations, i.e.
$\bfm{R}_{n,\perp}(z)=\bfm{R}_{n}^o+\bfm{u}_n(z)$. Up to second order
in the displacements ${\bfm u}_n(z)$, the energy cost can then be
expressed as
\begin{equation}
  \label{eq:g12}
  {\cal H}={\cal H}^o+\Delta {\cal H}.
\end{equation}
The first term ${\cal H}^o$, gives the free energy of an
array of $N$ straight flux lines oriented parallel to the external
field, and located at positions ${\bfm R}_n^o$, 
\begin{eqnarray}
  \label{eq:g13}
{\cal H}^o&=&L\frac{\phi_o^2}{8\pi}\sum_{n,m=1}^N\int \frac{
d^{2}\bbox{q}}{(2\pi)^2} \frac{{\rm e}^{i{\bfm q}\cdot
  {\bfm d}_{nm}}}{(\lambda^2 q^2 +1)} - L\frac{\phi_o N H}{4\pi} \nonumber \\
&=& N L\frac{\phi_o}{4\pi}(H_{c_1}-H)+L \epsilon_o \sum_{n,m\neq n}
 K_o\left(\frac{d_{nm}}{\lambda}\right).
\end{eqnarray}
Here $L$ is the sample thickness, $H_{c_1}=\phi_o/(4\pi\lambda^2)
\ln(\kappa)$ is the lower critical field, $\kappa=\lambda/\xi$ is the
Ginzburg-Landau parameter, $H$ is the magnetic field strength,
$\epsilon_o=\phi_o^2/(4\pi\lambda)^2$ is an interaction energy per
unit length, $K_o(x)$ is the modified Bessel function of zeroth order,
and ${\bfm d}_{nm}={\bfm R}_n^o-{\bfm R}_m^o$ are the relative
position vectors of any pair of flux lines.  On a perfect triangular
lattice, these vectors are of the form ${\bfm d}_{nm}=a
(n\hat{e}_1+m\hat{e}_2)$, ($n,m=0,\pm 1,\pm 2,\ldots$), with $a$ the
lattice spacing, $\hat{e}_1$ oriented, for instance, along the $x$
axis, $\hat{e}_1=\hat{x}$, and
$\hat{e}_2=\cos(\pi/3)\hat{x}+\sin(\pi/3)\hat{y}$ (see
Fig.~\ref{lattice}). The modulus of one of these vectors is then given
by $d_{nm}=a(n^2+m^2+nm)^{1/2}$.

The first term on the right hand side of \equ{eq:g13} represents the
energy cost of a single vortex line in a type II superconductor, times
the number of lines $N$. The last term is due to the interactions among
flux lines, which naturally depend on the interline separations.  
The penetration length $\lambda$, sets the extent of the interaction
potential $K_o(x)$, which diverges logarithmically at short distances,
and decays exponentially for $x\gg 1$.

The quantity $\Delta {\cal H}$ represents the {\em harmonic}
contribution of fluctuations to the free energy. After writing the
displacement fields ${\bfm u}_n(z)$ in terms of Fourier modes

\begin{equation}
  \label{eq:g14}
{\bfm u}_n(z)=\int \frac{dk_z}{(2\pi)}\int_{BZ} \frac{d^2{\bfm Q}}{(2\pi)^2}
 {\rm e}^{i({\bfm Q}\cdot {\bfm R}_n^o+k_zz)} {\bfm u}({\bfm Q},k_z),
\end{equation}
and taking advantage of the translational symmetry of the lattice, the
energy $\Delta {\cal H}$ can be written as
 
\begin{eqnarray}
  \label{eq:g15}
\Delta{\cal H}=\int
\frac{dk_z}{(2\pi)}\int_{BZ}\frac{d^{2}{\bfm Q}}{(2\pi)^2}&\mbox{\ }&u^{\alpha}({\bfm
Q},k_z) M_{\alpha\beta}({\bfm Q},k_z) \nonumber \\
&\mbox{\ }& \times\ u^{\beta}(-{\bfm Q},-k_z),
\end{eqnarray}
where the index $BZ$ indicates that the integration is performed over
the first Brillouin zone in reciprocal space. 

All the relevant information is therefore contained in the harmonic
kernel $M_{\alpha\beta}({\bfm Q},k_z)$.  Formally, we have to find the
eigenvalues and eigenvectors of this matrix, and we can then calculate
the extent of fluctuations in real space, the fluctuation corrections
to the free energy, and other relevant quantities.  In Fourier space,
the eigenvectors of $M_{\alpha\beta}({\bfm Q},k_z)$ are $N$
longitudinal modes, $u^L({\bfm Q},k_z)=\hat{Q}\cdot {\bfm u}({\bfm
Q},k_z)$, and $N$ transversal modes, $u^T({\bfm Q},k_z)=|\hat{Q}\times
{\bfm u}({\bfm Q},k_z)|$, with corresponding eigenvalues
$\Lambda_L({\bfm Q},k_z)$ and $\Lambda_T({\bfm Q},k_z)$.  In practice,
it turns out that the analytic expressions for these eigenvalues is
rather complex.  That is why in the literature only certain limiting
regimes are usually treated.  We shall describe these limits later on
in Sec.~\ref{sec:limits}, and then go on to discuss their accuracy in
comparison to our more general results.

\section{Longitudinal \& transversal modes}
\label{sec:moduli}

The longitudinal elastic eigenvalue $\Lambda_L({\bfm Q},k_z)$ is
typically expressed in the literature in terms of the so-called
compressional, $c_{11}(Q,k_z)$, and tilt, $c_{44}(Q,k_z)$, elastic
moduli as $c_{11}(Q,k_z) Q^2 + c_{44}(Q,k_z) k_z^2$.  (In general,
however, we shall see that this decomposition may be inaccurate.)  In
terms of the matrix $M_{\alpha\beta}({\bfm Q},k_z)$, the longitudinal
eigenvalue is given by

\begin{equation}
  \label{eq:m1}
  \Lambda_L({\bfm Q},k_z)= \hat{Q}_{\alpha}M_{\alpha \beta}({\bfm
  Q},k_z)\hat{Q}_{\beta}, 
\end{equation}
where, as usual, a repeated index is summed over.  
On the other hand, the transversal eigenvalue $\Lambda_T({\bfm Q},k_z)$,
commonly written in the literature in terms of the shear,
$c_{66}(Q,k_z)$, and tilt moduli as $c_{66}(Q,k_z) Q^2 + c_{44}(Q,k_z)
k_z^2$, is
\begin{equation}
  \label{eq:m2}
\Lambda_T({\bfm Q},k_z)=M_{\alpha\alpha}({\bfm Q},k_z)-
\hat{Q}_{\alpha}M_{\alpha\beta}({\bfm Q},k_z)\hat{Q}_{\beta}.
\end{equation}

\begin{figure}
\epsfxsize=4truecm{\centerline{\epsfbox{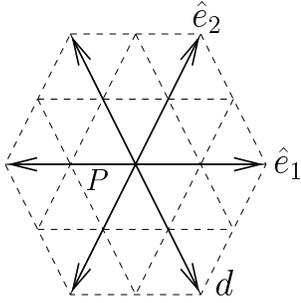}}}\medskip
\narrowtext{\caption{Lattice vectors with a common origin at the
    lattice point $P$, and with the same modulus $d$.\label{lattice} }}
\end{figure}

There are two alternative ways of expressing the interaction kernel
$M_{\alpha\beta}({\bfm Q},k_z)$: in terms of a sum over the Bravais
lattice vectors, or as a sum over the reciprocal lattice vectors. The
former yields
\begin{eqnarray}
  \label{eq:m3}
& & M_{\alpha\beta}({\bfm
   Q},k_z)=\frac{n\epsilon_o}{\lambda^2}~\Big\{\frac{1}{2}{\cal 
   E}(k_z)\delta_{\alpha\beta} +\nonumber \\
& &\sum_{m\neq n}[\cos({\bfm Q}\cdot{\bfm d}_{nm})
      R_{\alpha\beta}({\bfm d}_{nm},k_z)-R_{\alpha\beta}({\bfm
        d}_{nm},0)]\Big\},
\end{eqnarray}
where we have introduced the areal density of flux lines $n=N/A$. For
compactness of notation, let us also introduce the dimensionless
variables $x=\sqrt{\lambda^2 k_z^2 +1}$, and $x_c=\epsilon\ x$. In
\equ{eq:m3}, the quantity $n\epsilon_o~{\cal E}(k_z)/\lambda^2$
represents the line tension of each vortex, which, in general,
contains contributions from both the Josephson coupling between the
different Cu-O layers in the material, and the magnetic interlayer
interactions, as
\begin{equation}
  \label{eq:m4}
  {\cal E}(k_z)=\frac{1}{2} \epsilon^2 \lambda^2
  k_z^2\ln\left(\frac{\kappa^2}{x_c^2}+1\right)+\ln(x). 
\end{equation}
The interaction kernel $R_{\alpha\beta}({\bfm d},k_z)$ is given by
\begin{eqnarray}
 \label{eq:m5}
& &R_{\alpha \beta}({\bfm d},k_z)=\frac{\lambda}{d~x}\
K_1\left(\frac{d~x}{\lambda}\right)\delta_{\alpha
  \beta}-K_2\left(\frac{d~x}{\lambda}\right)  
\frac{d_{\alpha}d_{\beta}}{d^2} \nonumber \\
& & \mbox{\ }+ \epsilon^2 \lambda^2 k_z^2 \left\{
\left[K_o\left(\frac{d~x_c}{\lambda}\right) 
+\frac{\lambda}{d~x_c}\ K_1\left(\frac{d~x_c}{\lambda}\right)
\right]\delta_{\alpha
  \beta}\right. \nonumber \\
& & \mbox{\ }\left. -K_2\left(\frac{d~x_c}{\lambda}\right)  
\frac{d_{\alpha}d_{\beta}}{d^2}\right\}.
\end{eqnarray}
Note that we use the symbol $d=|\bfm d|$ to indicate the distance between
a pair of lattice points. The sum in Eq.~(\ref{eq:m3}) runs over all
such separations from a specific point on the lattice, as illustrated
in Fig.~\ref{lattice}.

The matrix $R({\bfm d},k_z)$ depends on the different relative
position vectors ${\bfm d}={\bfm d}_{nm}$ of the perfect triangular
lattice.  When summing over all lattice vectors with a common origin
at a point $P$, we can take advantage of the lattice symmetries.  For
example, the sum over products of an odd number of vector components
vanishes, as in
\begin{equation}
  \label{eq:m6}
  \sum_{{\bfm d}\neq 0} d_{\alpha} =0, \quad {\rm and, }
  \quad \sum_{{\bfm d}\neq 0} d_{\alpha} d_{\beta} d_{\gamma}=0.
\end{equation}
On the other hand, the sums involving products of an even number of 
components are nonvanishing and constrained by the symmetries;
for example,
\begin{eqnarray}
  \label{eq:m7}
& & \sum_{{\bfm d}\neq 0} d_{\alpha} d_{\beta} =
  \delta_{\alpha \beta}\sum_{{\bfm d}\neq 0}\frac{d^2}{2}, \\ \nonumber
& & \sum_{{\bfm d}\neq 0} d_{\alpha} d_{\beta} d_{\gamma}  d_{\delta}
= (\delta_{\alpha \beta} \delta_{\gamma\delta}+
\delta_{\alpha\gamma}\delta_{\beta\delta}+
\delta_{\alpha\delta}\delta_{\beta\gamma})
\sum_{{\bfm d}\neq 0} \frac{d^4}{8}.
\end{eqnarray}
(Higher order terms have a more complex structure.)  The sums
$\sum_{{\bfm d}\neq 0}$ on the right hand side of Eqs.~(\ref{eq:m7}),
are equivalent to $\sum_{d\neq 0} g(d)$, with a degeneracy factor
$g(d)$, which counts the number of vectors whose modulus is $d$.

Equivalently, we can express the kernel as a sum over reciprocal
lattice vectors ${\bfm G}$. The same symmetries
(\ref{eq:m6})-(\ref{eq:m7}) are, of course, valid for these vectors,
which, according to our choice of ${\bfm d_{nm}}$, are given by ${\bfm
G}=4\pi/(\sqrt{3}a)(p\hat{g}_1+q\hat{g}_2)$, ($p,q=0,\pm 1,\pm
2,\ldots$), where $4\pi/\sqrt{3}a$ is the reciprocal lattice spacing,
and the unit vectors are
$\hat{g}_1=\sin(\pi/3)\hat{x}-\cos(\pi/3)\hat{y}$, and
$\hat{g}_2=\hat{y}$.  In terms of these vectors, the interaction
kernel reads
\begin{eqnarray}
  \label{eq:m8}
& & M_{\alpha\beta}({\bfm Q},k_z)=\frac{n^2\phi_o^2}{8\pi}\Big\{
\sum_{{\bfm G}} [G_{\alpha\beta}^S({\bfm G}-{\bfm
  Q},k_z)-G_{\alpha\beta}({\bfm G},0)] \nonumber \\ 
& & \mbox{\ }-\frac{1}{n}\int
\frac{d^2{\bfm q}}{(2\pi)^2}[G_{\alpha\beta}^S({\bfm q}-{\bfm
  Q},k_z)-G_{\alpha\beta}({\bfm q},k_z)]\Big\}. 
\end{eqnarray}
The symbol $^S$ indicates symmetrization with respect to ${\bfm Q}$,
i.e., $G_{\alpha\beta}^S({\bfm G}-{\bfm Q},k_z)=
[G_{\alpha\beta}({\bfm G}- {\bfm Q},k_z)+G_{\alpha\beta}({\bfm G}+
{\bfm Q},k_z)]/2$.  (The second term on the right-hand side of the
expression is included to properly account for the particular case
$n=m$ and $z=z'$ in \equ{eq:g11}.) In addition, from \equ{eq:g11} we
obtain
\begin{eqnarray}
 \label{eq:m9} G_{\alpha\beta}({\bfm
q},k_z)&=&\frac{k_z^2}{(\lambda_c^2 q^2 +\lambda^2 k_z^2
+1)}\left(\delta_{\alpha\beta}-\frac{q_{\alpha}
q_{\beta}}{q^2}\right)\nonumber \\ &+&\frac{k^2}{(\lambda^2 k^2
+1)}\frac{q_{\alpha} q_{\beta}}{q^2}.
\end{eqnarray}

Our goal is to provide accurate values of the harmonic eigenvalues as
a function of ${\bfm Q}$ and $k_z$ within the BZ, and for different
concentrations of flux lines.  The dimensionless parameter $a/\lambda$
will be used as the indicator of the areal density of flux lines.  A
small value of $a/\lambda$, corresponds to a highly dense regime with
strong and non-local interactions among the flux lines.  On the other
hand, for $a/\lambda\gg 1$ the concentration of flux lines is very
low, and interactions among them very weak.  In this dilute limit, the
elastic behavior of the lattice reflects to properties of a single
flux line.  We thus evaluate numerically the harmonic eigenvalues as a
function $a/\lambda$ throughout the BZ. The equivalence of expressions
(\ref{eq:m3}) and (\ref{eq:m8}) renders the use of one or the other a
matter of convenience.  For instance, for very high areal densities of
vortices $n$, it is common to disregard the discreteness of the
underlying arrangement and approximate the sums over lattice positions
by integrals.  This so-called continuum limit is very often used in
the literature \cite{Sudbo91a,Sudbo91b,Fisher-rev91,Sardella92}. The
elastic moduli which follow from this approximation can be read
directly from \equ{eq:m9}, when only the reciprocal lattice vector
${\bfm G}={\bfm 0}$ is taken into account.  We shall comment on the
accuracy of this limit in the following sections.

\section{Limiting regimes}
\label{sec:limits}

In this section, we first present some special cases previously
discussed in the literature. By comparing these limits to our
numerical results, one can then see their range of validity and the
accuracy of the approximations involved in their formulation.  We
shall also emphasize the roles played by the interline distance, the
penetration length $\lambda$, and the symmetries of the triangular
lattice.

\subsection{Single line}
\label{sec:single}

At very low areal densities, i.e. $a/\lambda\gg 1$, the kernel
$M_{\alpha\beta}({\bfm Q},k_z)$ should be just $N$ times the result
obtained for a single flux line. All the modified Bessel functions
appearing in Eq.~(\ref{eq:m5}) decay exponentially fast for large
values of their argument (which is proportional to $a/\lambda$), and
only weakly contribute to the final values of the harmonic
eigenvalues. In other words, in \equ{eq:m3} only the line tension part
${\cal E}(k_z)$ is important for large values of $a/\lambda$.  In
Sec.~\ref{sec:numeric}, we show that the single-line approximation
indeed provides the best description of the elastic properties of the
system at very low concentrations of flux lines.

\subsection{Elastic moduli}
 \label{sec:local}

The energy cost of long wavelength distortions is described by an
elastic theory, whose form is constrained by symmetries.  The
triangular lattice is isotropic, and governed by the compressional,
$c_{11}$, shear, $c_{66}$, and tilt, $c_{44}$, moduli.  In terms of
the Fourier modes, the elastic energy is
\begin{eqnarray}
  \label{eq:l3}
& &\Delta{\cal H}=\frac{1}{2}\int \frac{dk_z}{2\pi}\int_{BZ}\frac{d^2{\bfm
    Q}}{(2\pi)^2}\left[c_{44} k_z^2\ \delta_{\alpha \beta} +
c_{11}Q_{\alpha}Q_{\beta}\right. \nonumber \\
& &\mbox{\ }+\left. c_{66}(Q^2\delta_{\alpha \beta}-Q_{\alpha}Q_{\beta})\right]
    u_{\alpha}({\bfm Q},k_z) u_{\beta}(-{\bfm Q},-k_z),  
\end{eqnarray}
and the harmonic eigenvalues have the simple forms
\begin{eqnarray}
\label{eq:l1}
\Lambda_L^l({\bfm Q},k_z)&=&(c_{11} Q^2 + c_{44} k_z^2)/2 \nonumber \\
\Lambda_T^l({\bfm Q},k_z)&=&(c_{66} Q^2 + c_{44} k_z^2)/2.
\end{eqnarray}
Formally, \equ{eq:g15} together with (\ref{eq:m3}) or (\ref{eq:m8})
provide the harmonic eigenvalues throughout the Brillouin zone.
Naturally, the elastic limit is regained by expanding these results up
to second order in $\bfm Q$ and $k_z$.  Neglecting the higher order terms is
referred to as the {\it local} limit, as it is usually obtained by
including only short-range interactions.  Sometimes, {\it non-local}
elastic moduli are introduced which depend on $\bfm Q$.  This is not
always useful, as it constrains the form of the harmonic eigenvalues
as in Eqs.(\ref{eq:l1}), whereas symmetry allows higher order powers
of $\bfm Q$ to appear in other forms.

After expanding the cosine, and using the symmetry properties of a
triangular lattice in \equs{eq:m6}{eq:m7}, as well as certain
relationship among the modified Bessel functions \cite{Bessel}, we
arrive at
\begin{eqnarray}
  \label{eq:l2}
& &M_{\alpha \beta}({\bfm
   Q},k_z)=\frac{n\epsilon_o}{\lambda^2}~\left[\frac{1}{2}{\cal  
   E}^l(k_z)\delta_{\alpha\beta}\right. \nonumber \\
& & + \frac{(\lambda k_z)^2}{2}\sum_{{\bfm d}\neq 0}
\left[\epsilon^2 K_o\left(\frac{\epsilon d}{\lambda}\right)
+\frac{1}{2}\left(\frac{d}{\lambda}\right) 
K_1\left(\frac{d}{\lambda}\right)\right]\delta_{\alpha \beta} \nonumber \\
& &+ \frac{Q^2}{2}\sum_{{\bfm d}\neq
 0}\frac{d^2}{4}\left\{\left[K_o\left(\frac{d}{\lambda}\right)+\frac{1}{2}K_2
\left(\frac{d}{\lambda}\right)\right]\frac{Q_{\alpha}Q_{\beta}}{Q^2}\right.
\nonumber \\
& &\left. +\left.\left[K_o\left(\frac{d}{\lambda}\right)-\frac{1}{2}K_2
\left(\frac{d}{\lambda}\right)\right]\left(\delta_{\alpha \beta}-
\frac{Q_{\alpha}Q_{\beta}}{Q^2}\right)\right\}\right]. 
\end{eqnarray}
where ${\cal E}^l(k_z)$ is the single line tension in the local
approximation, ${\cal E}^l(k_z)=(\epsilon \lambda k_z)^2
\ln(\kappa/\epsilon) + (\lambda k_z)^2/2$ (valid for
$\kappa\gg1\gg\epsilon$).

Comparing this expression with Eq.(\ref{eq:l3}) allows us to
identify the compressional modulus
\begin{equation}
  \label{eq:l4}
  c_{11}=\frac{n\epsilon_o}{4}\sum_{{\bfm d}\neq
 0}F_c\left(\frac{d}{\lambda}\right),
\end{equation}
with
\begin{equation}
  \label{eq:l5}
F_c\left(\frac{d}{\lambda}\right)=\frac{d^2}{\lambda^2}\left[K_o
\left(\frac{d}{\lambda}\right)+\frac{1}{2}K_2\left(\frac{d}{\lambda}\right)
\right],
\end{equation}
the shear modulus

\begin{equation}
  \label{eq:l6}
  c_{66}=\frac{n\epsilon_o}{4}\sum_{{\bfm d}\neq
 0}F_s\left(\frac{d}{\lambda}\right),
\end{equation}
with
\begin{equation}
  \label{eq:l7}
F_s\left(\frac{d}{\lambda}\right)=\frac{d^2}{\lambda^2}\left[K_o
\left(\frac{d}{\lambda}\right)-\frac{1}{2}K_2
\left(\frac{d}{\lambda}\right)\right],
\end{equation}
and the tilt modulus
\begin{equation}
  \label{eq:l8}
  c_{44}=n\epsilon_o\left[\epsilon^2\ln\left(\frac{\kappa}{\epsilon}\right)+\frac{1}{2}+\sum_{{\bfm d}\neq 
 0}F_t\left(\frac{d}{\lambda}\right)\right],
\end{equation}
with
\begin{equation}
  \label{eq:l9}
F_t\left(\frac{d}{\lambda}\right)=\left[\epsilon^2 K_o
\left(\frac{\epsilon d}{\lambda}\right) +
\frac{1}{2}\left(\frac{d}{\lambda}\right) K_1
\left(\frac{d}{\lambda}\right)\right].
\end{equation}
Note that the mean-field lattice spacing $a$, is related to the
strength of the applied field $H$ through the relationship
\begin{equation}
  \label{eq:l9b}
  H=H_{c_1}+\frac{\phi_o}{4\pi\lambda^2}\sum_{{\bfm d}\neq 0}\left[K_o
    \left(\frac{d}{\lambda}\right) +
    \frac{1}{2}\left(\frac{d}{\lambda}\right) K_1
    \left(\frac{d}{\lambda}\right)\right],
\end{equation}
which is obtained from $\partial H^o/\partial a|_a=0$. Then, by
comparison, the local tilt modulus for an isotropic material
($\epsilon=1$) can be written as $c_{44}=n\phi_o H/4\pi$, in agreement
with the result expected for the local tilt modulus of a rotationally
invariant superconductor \cite{Nelson89}. Strictly speaking, in
Eq.~(\ref{eq:l8}) there is an extra factor of $1/2$, coming from the
local line tension. However, as pointed out by Fisher
\cite{Fisher-rev91}, one could chose a short distance cutoff $\xi$
\cite{cutoff} inside $\ln(\lambda/\xi)$ to reproduce exactly the local
isotropic limit.

In Figs.~\ref{compshear}a) and \ref{compshear}b), we depict the
functions $F_c$ and $F_s$, for different values of the dimensionless
quantity $d/\lambda$, to illustrate some subtleties associated with
the estimation of the shear modulus, often forgotten or misunderstood
in the literature.  As shown in Fig.~\ref{compshear}b), for small
values of $d/\lambda$ the function $F_s$ is negative, whereas if
$d/\lambda>2$ it becomes positive and then decays exponentially to
zero. The implications of this functional form for the shear modulus
are as follows: As indicated before, the range of interactions among
vortex lines is determined by the penetration length $\lambda$.  For a
dilute vortex lattice whose lattice spacing $a$ is comparable or even
greater than $\lambda$, almost all the terms in the sum in
Eq.~(\ref{eq:l6}) are positive, and as they rapidly decay to zero,
only the first few lattice vectors for neighboring sites are needed to
calculate the shear modulus. On the other hand, in a dense system
there are many neighboring vortices within the interaction range
$\lambda$, contributing a {\it negative} amount to the sum in
\equ{eq:l6}.  It is then clearly not sufficient to consider only
interactions among a few neighboring lines, and to account for the
stability of a dense lattice against shear deformations, we need to
sum over many lattice vectors (finally giving rise to a positive shear
modulus).

\end{multicols}\widetext

\renewcommand{\caption}[1]{\refstepcounter{figure}\protect\noindent%
  \protect\parbox{17.8cm}{\small FIG. \thefigure. #1}}

\begin{figure}
\centerline{\hbox{\epsfysize=8truecm\epsfbox{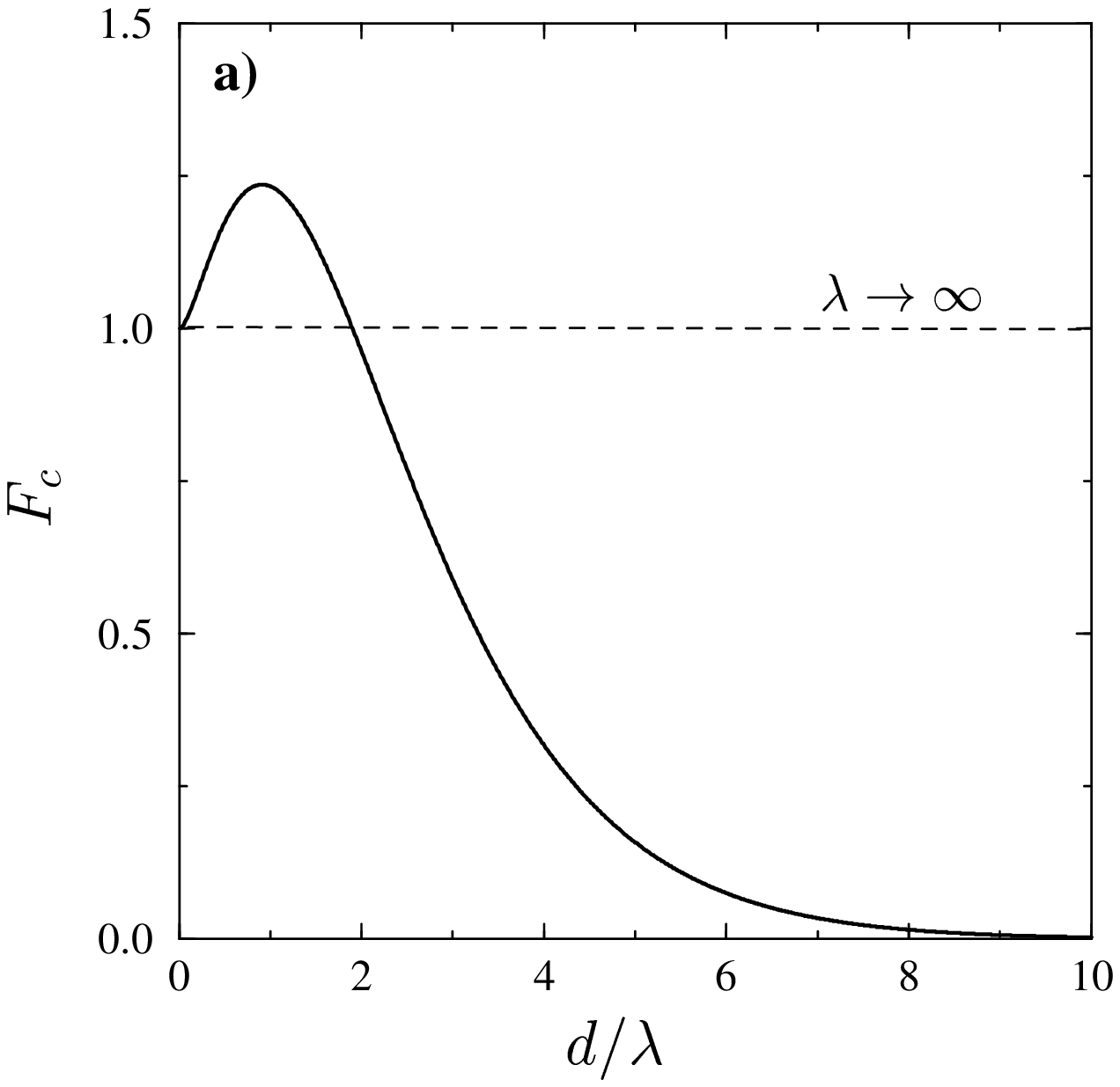}\
\epsfysize=8truecm\epsfbox{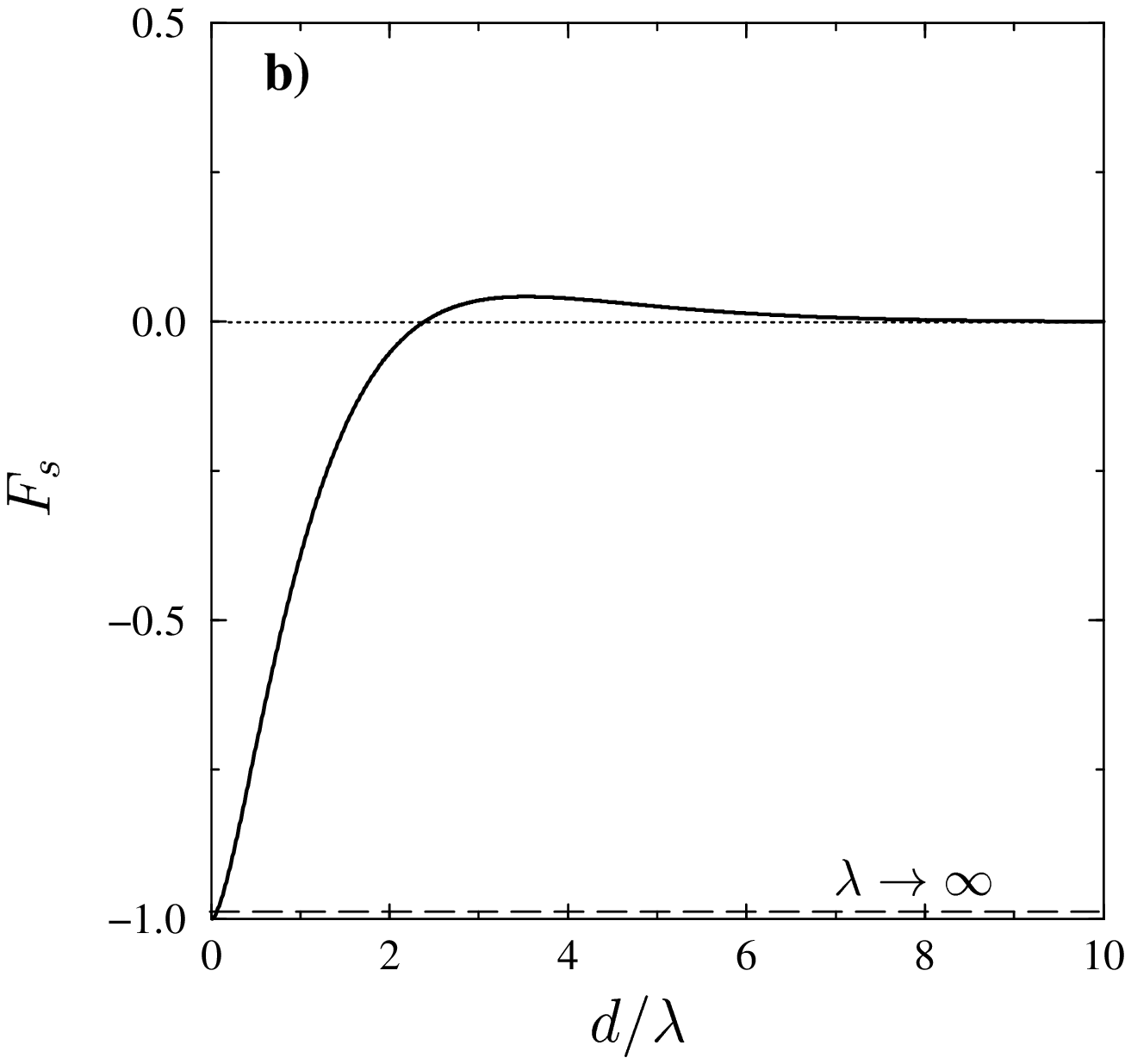}}}
\medskip
\caption{a) Behavior of $F_c$ in \equ{eq:l5} as a function of the
dimensionless parameter $d/\lambda$. b) Behavior of the function $F_s$
in \equ{eq:l7} as a function of the dimensionless parameter
$d/\lambda$. The long-dashed lines represent the $\lambda\rightarrow
\infty$ limit of these functions.\label{compshear}}
\end{figure} 


\begin{multicols}{2}\narrowtext

In Figs.~\ref{compshear}a) and \ref{compshear}b), we also indicate
with a dashed line the limit $\lambda\rightarrow \infty$ (for a finite
value of $d$).  In this case there is a logarithmic interaction among
the vortices, as in the two dimensional Coulomb gas \cite{TH}. Note
that in this limit, all factors in the sum for the shear modulus are
negative, and the lattice is unstable to shear deformation.  (This is
a manifestation of the more general instability of a system of point
charges in the absence of external potentials.)

\begin{figure}
\centerline{
\hbox{\epsfysize=8truecm\epsfbox{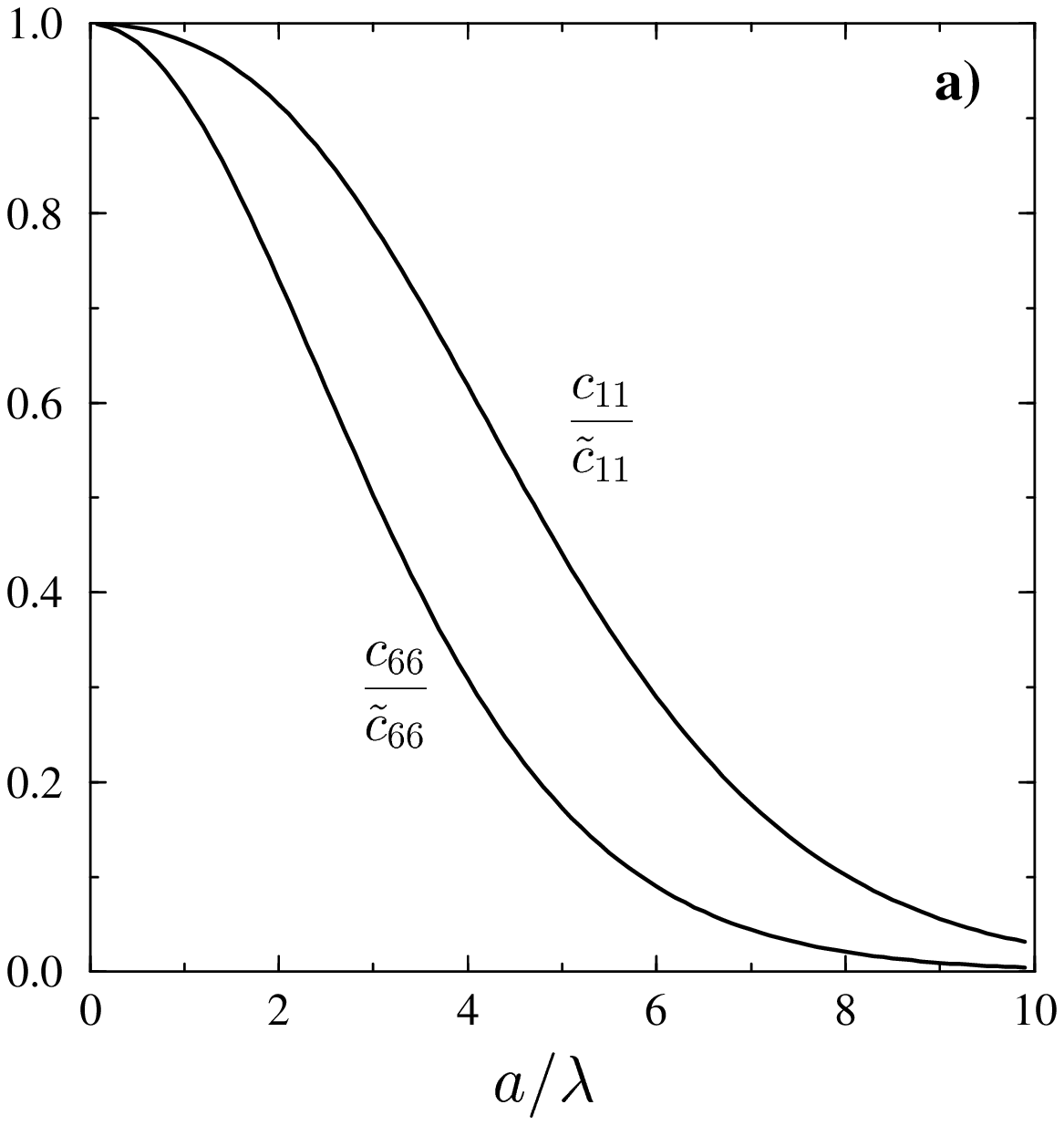}\
\epsfysize=8truecm\epsfbox{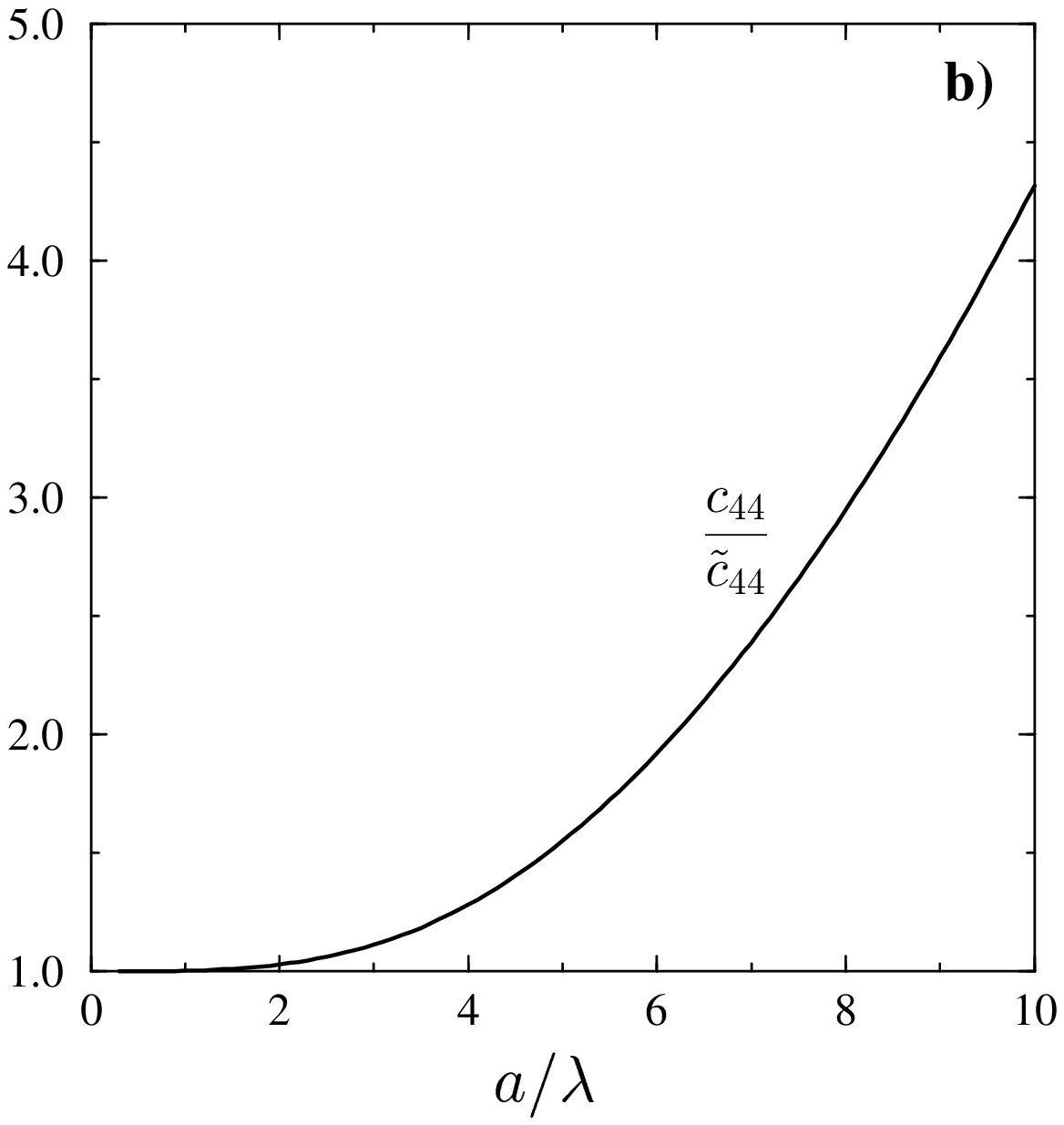}}}
\medskip
\caption{a) Normalized compressional and shear moduli as a function of
the dimensionless parameter $a/\lambda$, obtained by numerical
evaluation of the sums in Eqs.~(\ref{eq:l4}) and (\ref{eq:l6}). b)
Normalized tilt modulus as a function of the dimensionless parameter
$a/\lambda$, from numerical evaluation of the sum in Eq.(\ref{eq:l8}).
\label{compressheartilt}} 
\end{figure}

Figure~\ref{compressheartilt} depicts the compressional, shear, and
tilt moduli as a function of $a/\lambda$, obtained by numerical
evaluation of the sums in Eqs.(\ref{eq:l4}), (\ref{eq:l6}), and
(\ref{eq:l8}), respectively.  These moduli have been normalized by
their respective values in the {\em local continuum limit}, which for
the long wavelength compressional modulus is defined as

\begin{eqnarray}
  \label{eq:l10}
 \tilde{c}_{11}&=&\frac{n\epsilon_o}{4}\left\{-F_c(d\rightarrow
  0)+\int \frac{d^2{\bfm d}}{{\cal
  A}_{pc}}F_c\left(\frac{d}{\lambda}\right)\right\} \nonumber \\ 
&=&\frac{n\epsilon_o}{4}\left(-1+\frac{16\pi\lambda^2}{{\cal
  A}_{pc}}\right)=\frac{B^2}{4\pi}-\frac{B\phi_o}{(8\pi\lambda)^2}, 
\end{eqnarray}
where we have replaced the sums by integrals, after first adding and
susbtracting the ${\bfm d}\rightarrow {\bfm 0}$ element explicitly,
and introducing the appropriate unit area ${\cal
A}_{pc}=\sqrt{3}a^2/2$, i.e. the area of the {\em primitive cell}.  We
have also defined the average magnetic induction $B=n\phi_o$, with
$n=N/A=1/{\cal A}_{pc}$.  Similarly, the long wavelength shear
modulus is

\begin{eqnarray}
  \label{eq:l11}
 \tilde{c}_{66}&=&\frac{n\epsilon_o}{4}\left\{-F_s(d\rightarrow
  0)+\int \frac{d^2{\bfm d}}{{\cal A}_{pc}}
  F_s\left(\frac{d}{\lambda}\right)\right\} \nonumber \\ 
&=&\frac{n\epsilon_o}{4}(1+0)=\frac{B\phi_o}{(8\pi\lambda)^2}, 
\end{eqnarray}
and, for the tilt modulus we obtain

\begin{equation}
  \label{eq:l12}
 \tilde{c}_{44}= n\epsilon_o \int \frac{d^2{\bfm d}}{{\cal A}_{pc}}
  F_t\left(\frac{d}{\lambda}\right)  
= n \epsilon_o \frac{4\pi\lambda^2}{{\cal A}_{pc}}=\frac{B^2}{4\pi}.
\end{equation}
Note that for the latter, the term ${\bfm d}\rightarrow {\bfm 0}$ is
already included in \equ{eq:l8}. 

Similar results for the continuum local limit have been previously
reported in the literature \cite{Blatter94,Brandt77,Kogan89}. In
Ref.~\onlinecite{Kogan89}, Kogan and Campbell calculated the shear
modulus of a flux lattice in different situations. Essentially, they
provided an expression for the shear modulus $c_{66}$ in terms of a
sum over components of the reciprocal lattice vectors.  Because of the
symmetry relations of Eq.(\ref{eq:m7}) quoted in the previous section, all
the terms included in their sums cancel each other out, and should
yield a shear modulus equal to zero.  Kogan and Campbell numerically
evaluated this sum over reciprocal lattice vectors, and reported a
value close to Eq.(\ref{eq:l11}). They argued that there is a
contribution due to the specific form of the cut-off at short
distances \cite{cutoff}.  However, in view of the above derivation of
Eq.(\ref{eq:l11}), it is clear that cutoffs are not relevant  to the
local continuum value of the shear modulus.

As soon as the lattice spacing is greater than the penetration length,
$a>\lambda$, all the compressional, shear, and tilt moduli deviate
strongly from the continuum results. The compressional and shear
moduli then decay exponentially to zero, while the tilt modulus
attains a constant value of $n\epsilon_o
(\epsilon^2\ln(\kappa/\epsilon)+1/2)$.  (This is why when normalized
by its continuum value of Eq.(\ref{eq:l12}) the ratio grows as
$(a/\lambda)^2$.)  As we shall demonstrate in Sec.(\ref{sec:melting}),
under these circumstances the lattice again becomes rather soft and
liable to melting, as its fluctuations are greatly enhanced for large
interline separations.

\end{multicols}\widetext

\begin{multicols}{2}\narrowtext

\subsection{(Non-local) Continuum limit}
\label{sec:continuum}

The continuum limit refers to a situation in which the concentration
of flux lines is high. In this limit, one disregards the underlying
triangular lattice arrangement and treats the array of vortices as a
continuous medium \cite{Sudbo91a,Sudbo91b,Fisher-rev91,Sardella92}. In
the continuum limit, the sum of Eq.~(\ref{eq:m3}) over Bravais lattice
vectors is approximated by an integral, or equivalently, in the sum
over reciprocal lattice vectors in Eq.~(\ref{eq:m8}) only the vector
$\bbox{G=0}$ is considered. As a result, the elastic kernel reduces to

\begin{eqnarray}
  \label{eq:lc1}
& & M_{\alpha\beta}({\bfm Q},k_z)=\frac{n^2\phi_o^2}{8\pi}\left\{
 \frac{(Q^2+k_z^2)}{[\lambda^2 (Q^2+k_z^2)
    +1]}\frac{Q_{\alpha} Q_{\beta}}{Q^2} \right.\nonumber \\
  & & + \frac{k_z^2}{(\lambda_c^2 Q^2 +\lambda^2 k_z^2 +1)} 
\left(\delta_{\alpha\beta}-\frac{Q_{\alpha}
        Q_{\beta}}{Q^2}\right)  \nonumber \\
  & & -\left. \frac{1}{n}\int \frac{d^2{\bfm
        q}}{(2\pi)^2}[G_{\alpha\beta}^S({\bfm q}-{\bfm
      Q},k_z)-G_{\alpha\beta}({\bfm q},k_z)]\right\}.
\end{eqnarray}
The last term in this equation is a rather intricate integral, which,
on the other hand, can be easily evaluated in the local approximation,
i.e. when keeping terms up to $Q^2$ and $k_z^2$ only. 
With this approximation, we can rewrite Eq.~(\ref{eq:lc1}) as

\begin{eqnarray}
  \label{eq:lc2} & & M_{\alpha\beta}({\bfm Q},k_z) \simeq
\frac{n^2\phi_o^2}{8\pi} \left\{\frac{(Q^2+k_z^2)}{[\lambda^2
(Q^2+k_z^2)+1]} \frac{Q_{\alpha}Q_{\beta}}{Q^2}\right.  \nonumber \\ &
& + \frac{k_z^2}{(\lambda_c^2 Q^2 +\lambda^2 k_z^2+1)}
\left(\delta_{\alpha\beta}-\frac{Q_{\alpha}
Q_{\beta}}{Q^2}\right)\nonumber \\ & & + \left.
\frac{1}{n}\left[\frac{Q^2\delta_{\alpha\beta}}{16\pi\lambda^2}-
\frac{Q_{\alpha} Q_{\beta}}{8\pi\lambda^2} \right]\right\}.
\end{eqnarray}
By comparing with Eq.~(\ref{eq:l3}), we can now directly read off
the nonlocal continuum (again, up to the approximation carried out
to evaluate the integral) values of the elastic moduli as

\begin{equation}
  \label{eq:lc4}
  c_{44}^{nl}=\frac{B^2}{4\pi}
    \frac{1}{(\lambda_c^2 Q^2 +\lambda^2 k_z^2+1)},
\end{equation} 

\begin{equation}
  \label{eq:lc5} c_{11}^{nl}=\frac{B^2}{4\pi}
  \frac{[\lambda_c^2(Q^2+k_z^2)+1]}{[\lambda^2
  (Q^2+k_z^2)+1](\lambda_c^2 Q^2 +\lambda^2 k_z^2 +1)}
  -\frac{B\phi_o}{(8\pi\lambda)^2},
\end{equation} 

\begin{equation}
  \label{eq:lc6}
  \tilde{c}_{66}=\frac{B\phi_o}{(8\pi\lambda)^2}.
\end{equation} 
Note that, up to second order in $Q^2$ and $k_z^2$,
Eqs.~(\ref{eq:lc4}) through (\ref{eq:lc6}) reproduce the local limits
calculated in the previous section from the real space expression of
the interaction kernel.

\section{Numerical results}
\label{sec:numeric}

In this section, we present and discuss our results from numerical
evaluations of \equs{eq:m1}{eq:m2} using, in particular, the real
space expression of the interaction matrix in Eq.~(\ref{eq:m3}).  In
most cases, we selected a Ginzburg-Landau parameter $\kappa$ equal to
$100$, and an anisotropy ratio of $\epsilon=0.1$.  However, later in
this section, we present some results for different values of
$\epsilon$, which may be more appropriate to describe materials with
stronger anisotropy.  For the sake of simplicity, we introduce the
dimensionless eigenvalues
$\Lambda_L^*=\Lambda_L/(n\epsilon_o/\lambda^2)$ and
$\Lambda_T^*=\Lambda_T/(n\epsilon_o/\lambda^2)$, and discuss their
behavior throughout this section.

\renewcommand{\caption}[1]{\refstepcounter{figure}\protect\noindent%
  \protect\parbox{8.6cm}{\small FIG. \thefigure. #1}}

\begin{figure}
\epsfxsize=8truecm{\centerline{\epsfbox{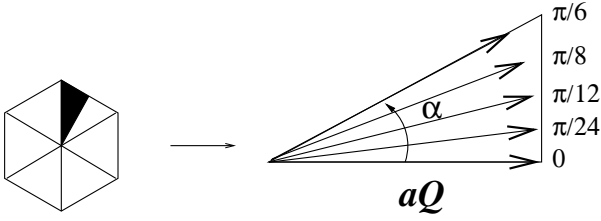}}}\medskip
\narrowtext{\caption{Irreducible Brillouin zone of a regular vortex
lattice.\label{IBZ}}}
\end{figure} 


\subsection{Angular dependence}
\label{sec:angle}

Because of the point group symmetries of the triangular lattice, it is
sufficient to consider only the {\em irreducible} Brillouin zone
(IBZ), indicated in black in Fig.~\ref{IBZ}.  In particular, we have
chosen five paths corresponding to values of the angle $\alpha$
(defined within the figure) equal to $0,\pi/24,\pi/12,\pi/8,
\mbox{and\ } \pi/6$.  Figure \ref{angles} shows the numerically
evaluated eigenvalues along these paths within the IBZ. The horizontal
axis is the dimensionless quantity $(aQ)^2$.  The plots also
correspond to the choice of $\lambda k_z=1.0$.  The dotted lines
represent the longitudinal eigenvalue, $\Lambda_L^*$, while we use
dashed lines for the transversal eigenvalue, $\Lambda_T^*$.

The different plots in Fig.~\ref{angles}a) clearly show that there is
an angular dependence, which becomes more marked as we approach the
edge of the BZ. This fact was already pointed out by Brandt in his
early work on the elastic properties of vortex lattices
\cite{Brandt77}.  The anisotropy is more pronounced in the dense
regime, and diminishes as we increase $a/\lambda$.  In the
concentrated case with $a/\lambda=0.2$, the relative differences
$|\Lambda(\alpha)-\Lambda(\pi/12)|/\Lambda(\pi/12)$ are up to $20\%$
for the longitudinal eigenvalue, and $60\%$ for the transversal one.
These ratios decrease in the intermediate region with $a/\lambda=1.0$,
and are of the order of $10\%$ and $15\%$, respectively, for
$a/\lambda=5.0$, within the dilute regime.  We also observe that the
longitudinal eigenvalue is largest along the $\alpha=0$-direction,
while the transversal eigenvalue attains its smallest value for
$\alpha=0$.

Note that both eigenvalues, for all angles $\alpha$, have the same
value at $Q=0$, determined by $\lambda k_z$ and $a/\lambda$.  (Since
this is not immediately apparent from Fig.~\ref{angles}, the
intersection point is marked with a black dot on the vertical axis.)
Away from $Q=0$, the transversal eigenvalue drops sharply, more so in
the dense regime.  Transversal modes are therefore less costly than
longitudinal modes.  Figure~\ref{angles} also indicates that at a
finite value of $\lambda k_z$ (equal to 1.0 in this case), the minimum
cost is obtained for a finite value of the rescaled wavevector $aQ$,
which depends on the density of flux lines.

\end{multicols}\widetext

\renewcommand{\caption}[1]{\refstepcounter{figure}\protect\noindent%
  \protect\parbox{17.8cm}{\small FIG. \thefigure. #1}}

\begin{figure}
\centerline{\hbox{\epsfysize=6truecm\epsfbox{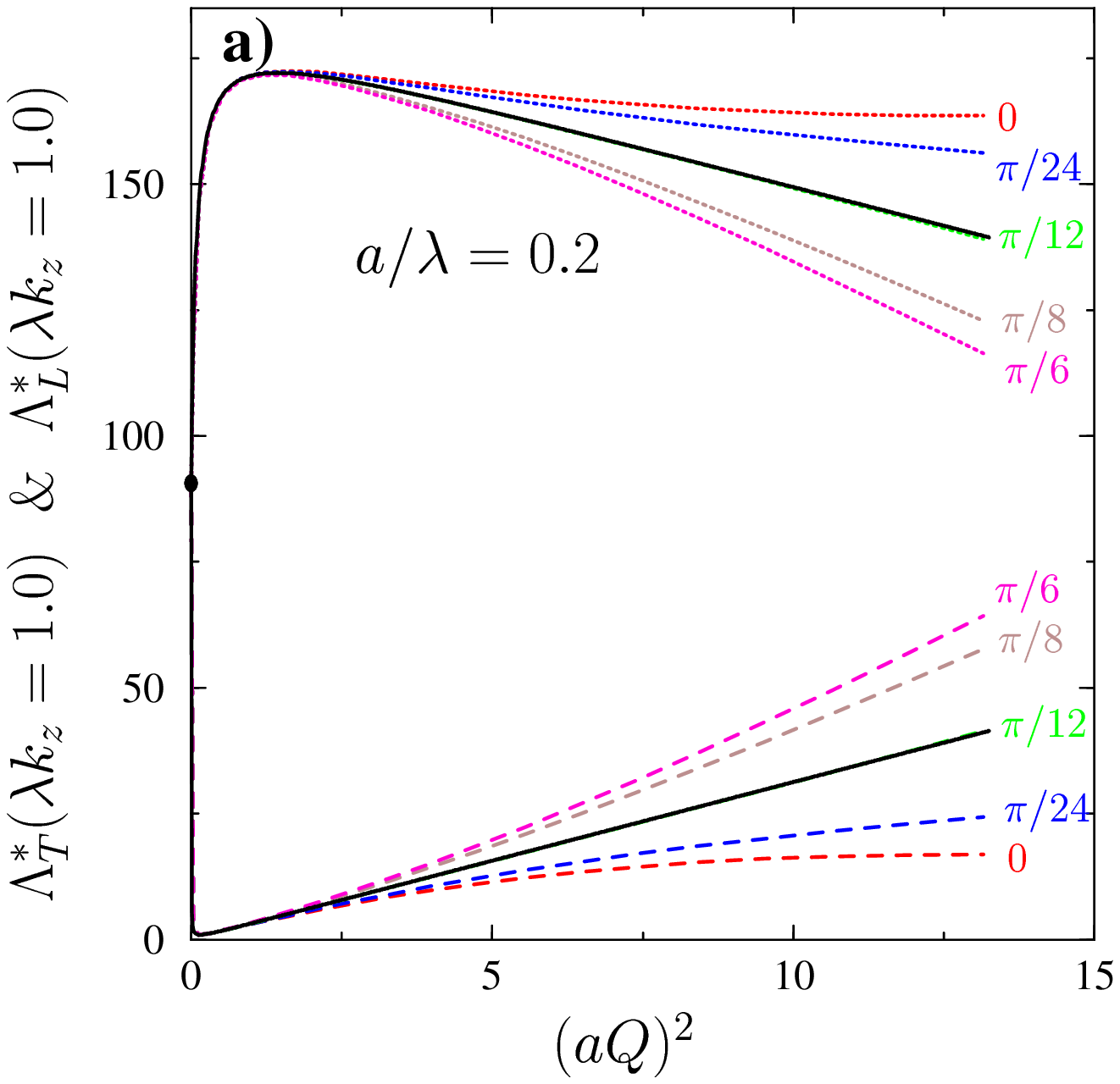}\
\epsfysize=6truecm\epsfbox{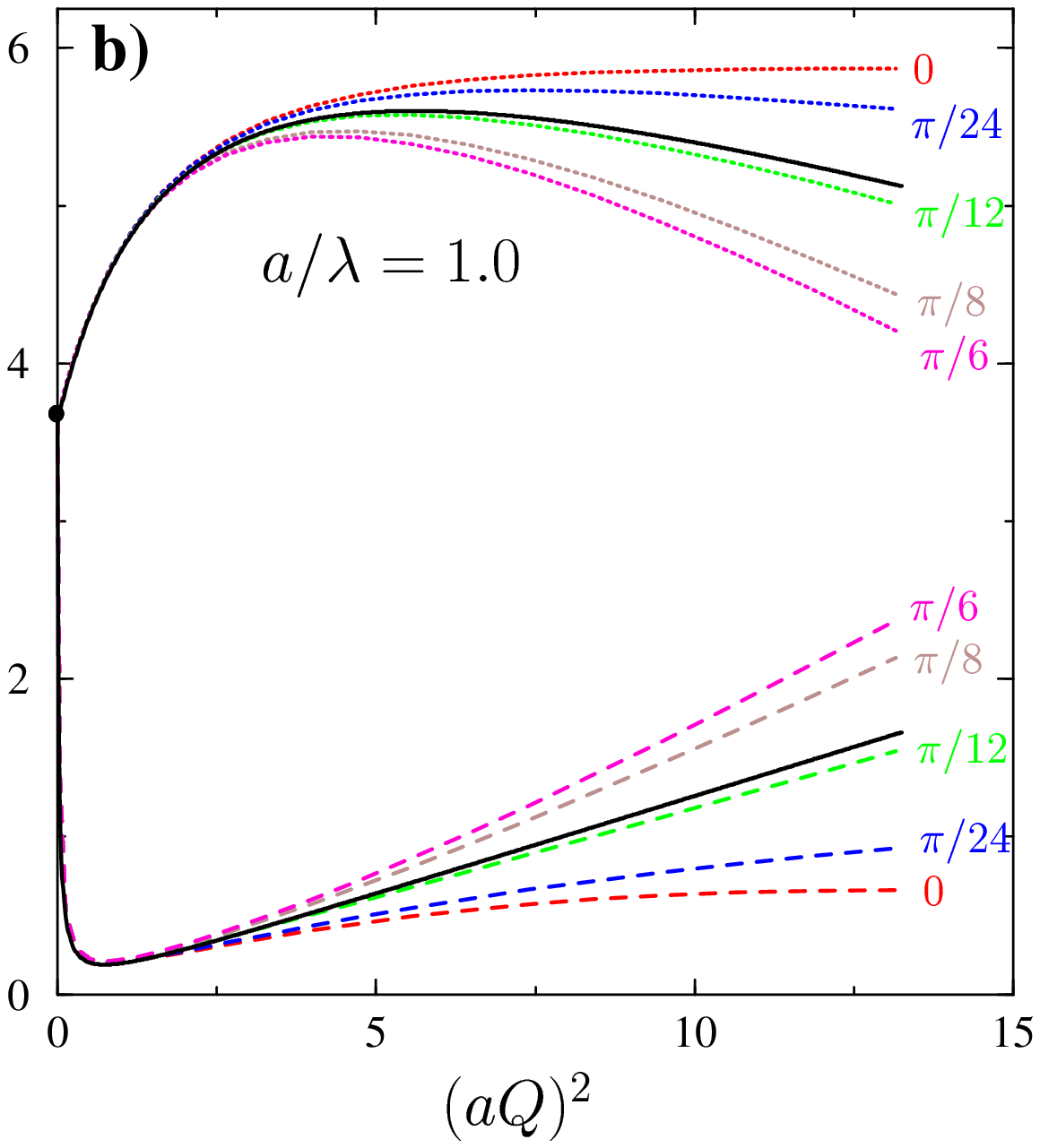}\ 
\epsfysize=6truecm\epsfbox{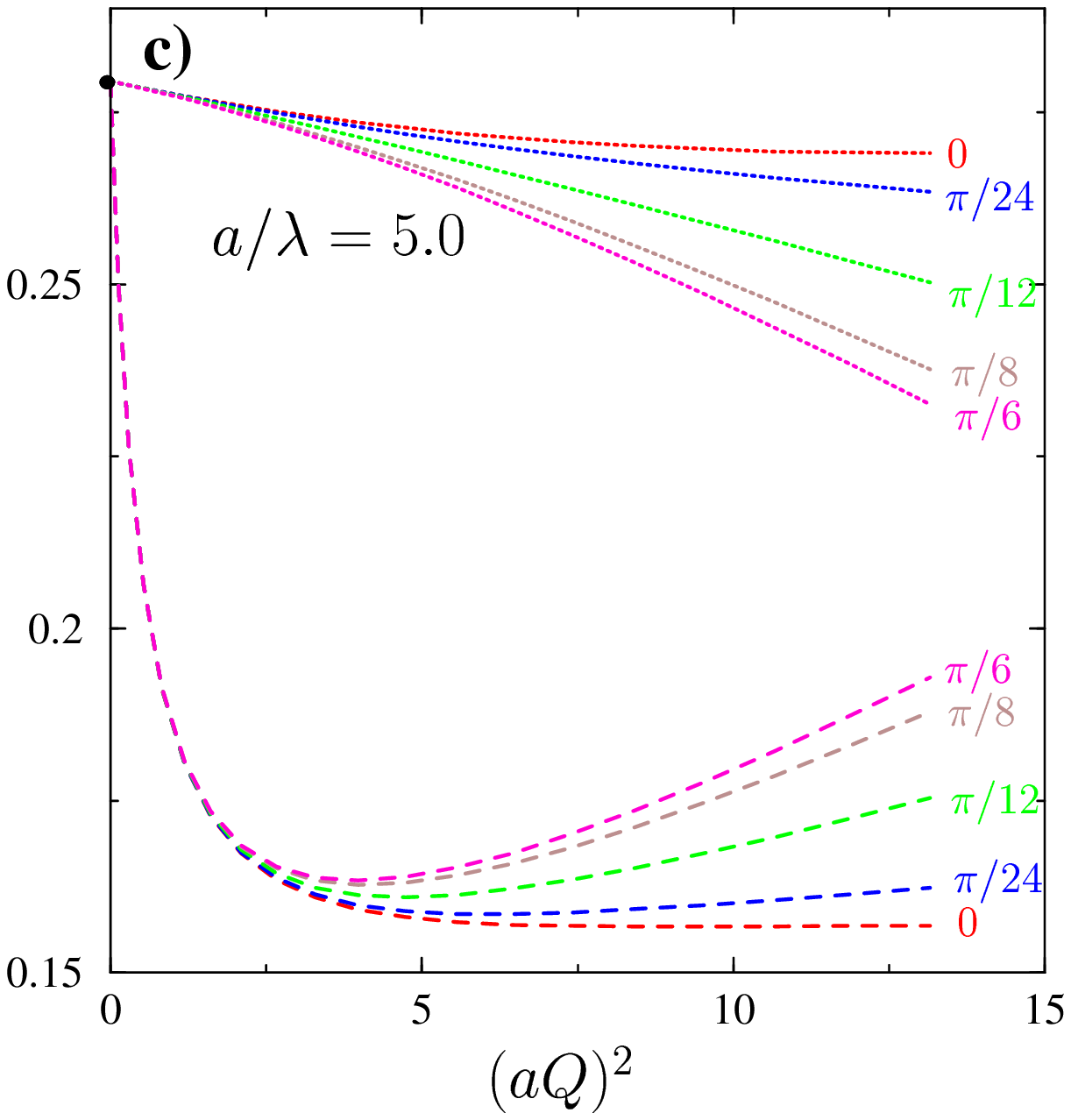}}}
\medskip
\caption{Dimensionless elastic eigenvalues $\Lambda_T^*$ and
$\Lambda_L^*$ as a function of $(aQ)^2$, for $\lambda k_z=1.0$, and at
a) $a/\lambda=0.2$, b) $a/\lambda=1.0$, and c) $a/\lambda=5.0$.  The
dotted lines correspond to the longitudinal eigenvalue, while dashed
lines indicate the transversal eigenvalue.  The different lines are
obtained for different paths along the IBZ, at angles
$\alpha=0,\pi/24,\pi/12,\pi/8, \mbox{and\ } \pi/6$. The two solid
lines in a) and b) depict eigenvalues from the nonlocal continuum
limit in Sec.~\ref{sec:limits}.  \label{angles} }
\end{figure} 


\begin{multicols}{2}\narrowtext

The solid lines in Figs.~\ref{angles}a) and \ref{angles}b) are the
results of the nonlocal continuum approximation introduced in
subsection \ref{sec:continuum}, i.e.,
$\Lambda_L\simeq(c_{11}^{nl}Q^2+c_{44}^{nl}k_z^2)/2$ and
$\Lambda_T\simeq(\tilde{c}_{66}Q^2+c_{44}^{nl}k_z^2)/2$, with the
appropriate parameters.  From these figures, one can conclude that (at
least for $\lambda k_z=1.0$) the continuum approximation reproduces
extremely well the behavior of both eigenvalues along the
$\alpha=\pi/12$ direction.  This is certainly the case at the highest
concentration of flux lines ($a/\lambda=0.2$), and even at the
intermediate concentrations of $a/\lambda=1.0$.  However, the
approximation fails at lower densities, and is not even included in
Fig.~\ref{angles}c), as the predicted eigenvalues fall well outside
the range of our plot.

\subsection{Variations with $k_z$}
\label{sec:kz}

To explore the $k_z$-dependence of the results, in Fig.~\ref{cskz} we
plot the longitudinal and transversal eigenvalues as a function of
$(aQ)^2$ for a fixed angle $\alpha=\pi/12$, and for a number of values
of $a/\lambda$ and $\lambda k_z$.  From these figures, we see that,
not surprisingly, both eigenvalues increase as a function of $\lambda
k_z$, and consequently, modes with higher $k_z$ are in general costly.
For the smallest value of $a/\lambda=0.2$ in Fig.~\ref{cskz}a), the
$\lambda k_z$ dependence is most pronounced in a small region close to
the BZ center. This region becomes larger as we increase either
$\lambda k_z$ or $a/\lambda$.  We also note that both eigenvalues show
a linear dependence on $(aQ)^2$ for a large fraction of the IBZ. The
slope of this line is almost constant within each plot, i.e. for a
given value of $a/\lambda$.  The transversal eigenvalue for $\lambda
k_z=0$ is a straight line throughout the whole range of values of
$(aQ)^2$, as is the longitudinal eigenvalue for high enough values of
$\lambda k_z$.  These features are qualitatively well accounted for by
the continuum limit results. According to Eqs.~(\ref{eq:lc4}),
(\ref{eq:lc5}), and (\ref{eq:lc6}), the transversal eigenvalue

\begin{eqnarray}
  \label{eq:n1}
 \Lambda_T(Q,k_z) &\simeq&
 \frac{1}{2}[\tilde{c}_{66}Q^2+c_{44}^{nl}(Q,k_z)k_z^2] \nonumber \\
&=& \frac{n\epsilon_o}{8}Q^2 + 2\pi n^2 \epsilon_o \frac{\lambda^2
 k_z^2}{(\lambda_c^2 Q^2+ \lambda^2 k_z^2+1)}
\end{eqnarray} 
reduces to $\tilde{c}_{66}Q^2/2$ for $\lambda k_z=0$, i.e., grows
linearly with $(aQ)^2$, with a slope $n\epsilon_o/8a^2$. In addition,
for suficiently large values of $\lambda k_z\gg \lambda_c Q$,
$c_{44}^{nl}(Q,k_z)k_z^2$ reaches a saturation value, rendering the
term $\tilde{c}_{66}Q^2/2$ as the leading contribution depending on
$Q$. Similarly, the longitudinal eigenvalue

\begin{eqnarray}
  \label{eq:n2}
 \Lambda_L(Q,k_z) &\simeq& \frac{1}{2}[c_{11}^{nl}(Q,k_z)Q^2+
c_{44}^{nl}(Q,k_z)k_z^2]   \nonumber \\
&=& 2\pi n^2 \epsilon_o \frac{\lambda^2(Q^2+k_z^2)}{[\lambda^2(Q^2+k_z^2)+1]} -\frac{n\epsilon_o}{8}Q^2
\end{eqnarray} 
reaches a saturation value for $\lambda k_z \gg \lambda Q$, so that
the term $-n\epsilon_o Q^2/8$ controls essentially the $Q$ dependence
of the eigenvalue at large values of $\lambda k_z$.
\equs{eq:n1}{eq:n2} fit very well the numerical resuls obtained for
$a/\lambda=0.2$ and $\lambda k_z=0.0,1.0,10.0$ (we are not showing
these results in Fig.~\ref{cskz}a) for the sake of clearity of the
figure).  For the larger value of $\lambda k_z=50.0$, although the
qualitative form of the functions is quite well approximated by these
equations, there are major quantitative differences with respect to
the numerical results mainly due to a smaller value of both
eigenvalues at $Q=0$. In Fig.~\ref{cskz}b) with $a/\lambda=1.0$, there
are important differences in the value at $Q=0$ already for $\lambda
k_z=10.0$, as well as small deviations from the value of the slope of
the linear part of the curves. These deficiencies of the continuum
approximation are even more evident for $a/\lambda=5.0$ in
Fig.~\ref{cskz}c).  In the latter case, we also note the emergence of
single-line characteristics: The eigenvalues are clearly mostly a
function of $k_z$, with only small variations with $(aQ)^2$.

\end{multicols}\widetext

\renewcommand{\caption}[1]{\refstepcounter{figure}\protect\noindent%
  \protect\parbox{17.8cm}{\small FIG. \thefigure. #1}}

\begin{figure}
\centerline{\hbox{\epsfysize=6truecm\epsfbox{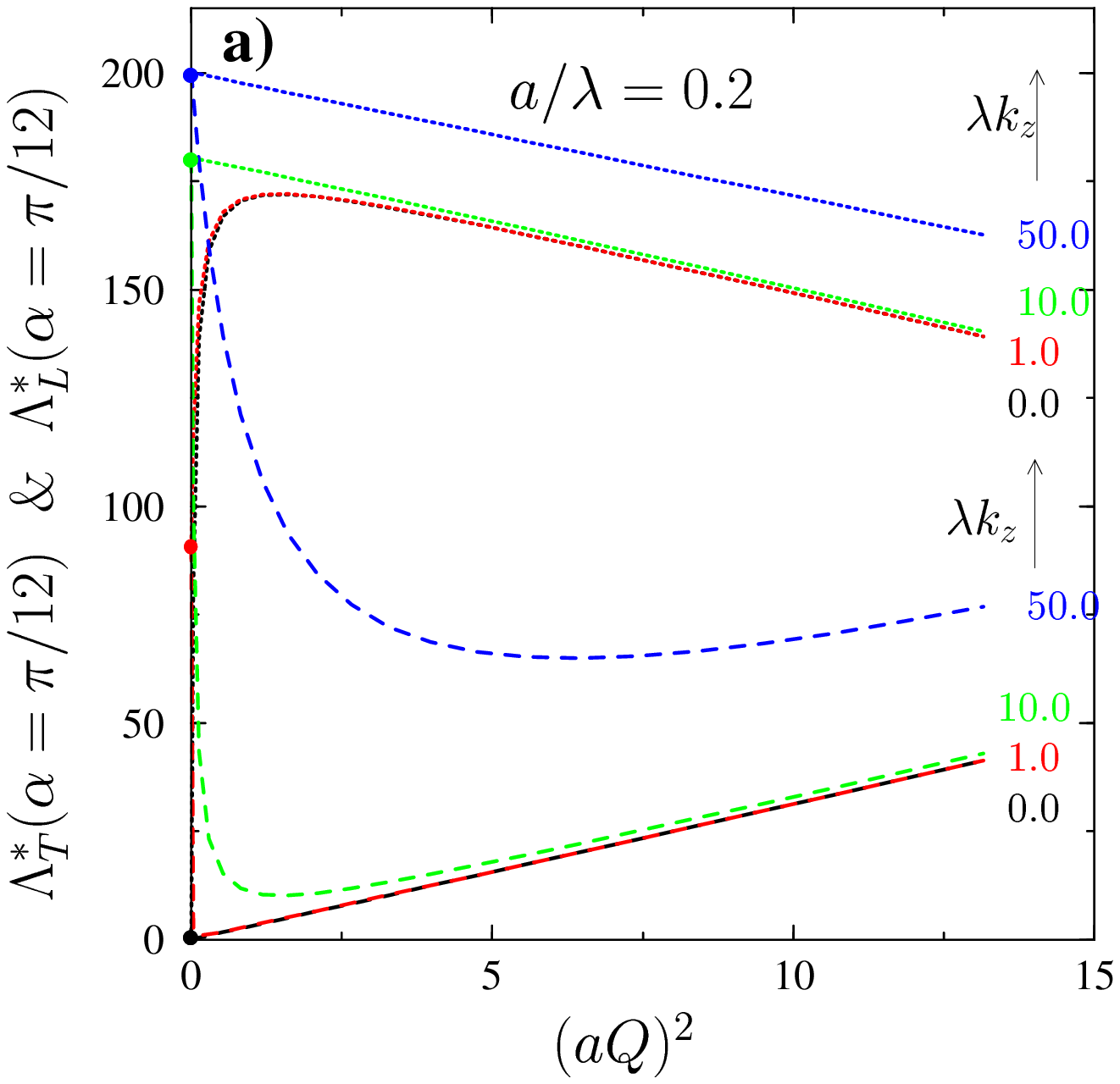}\
\epsfysize=6truecm\epsfbox{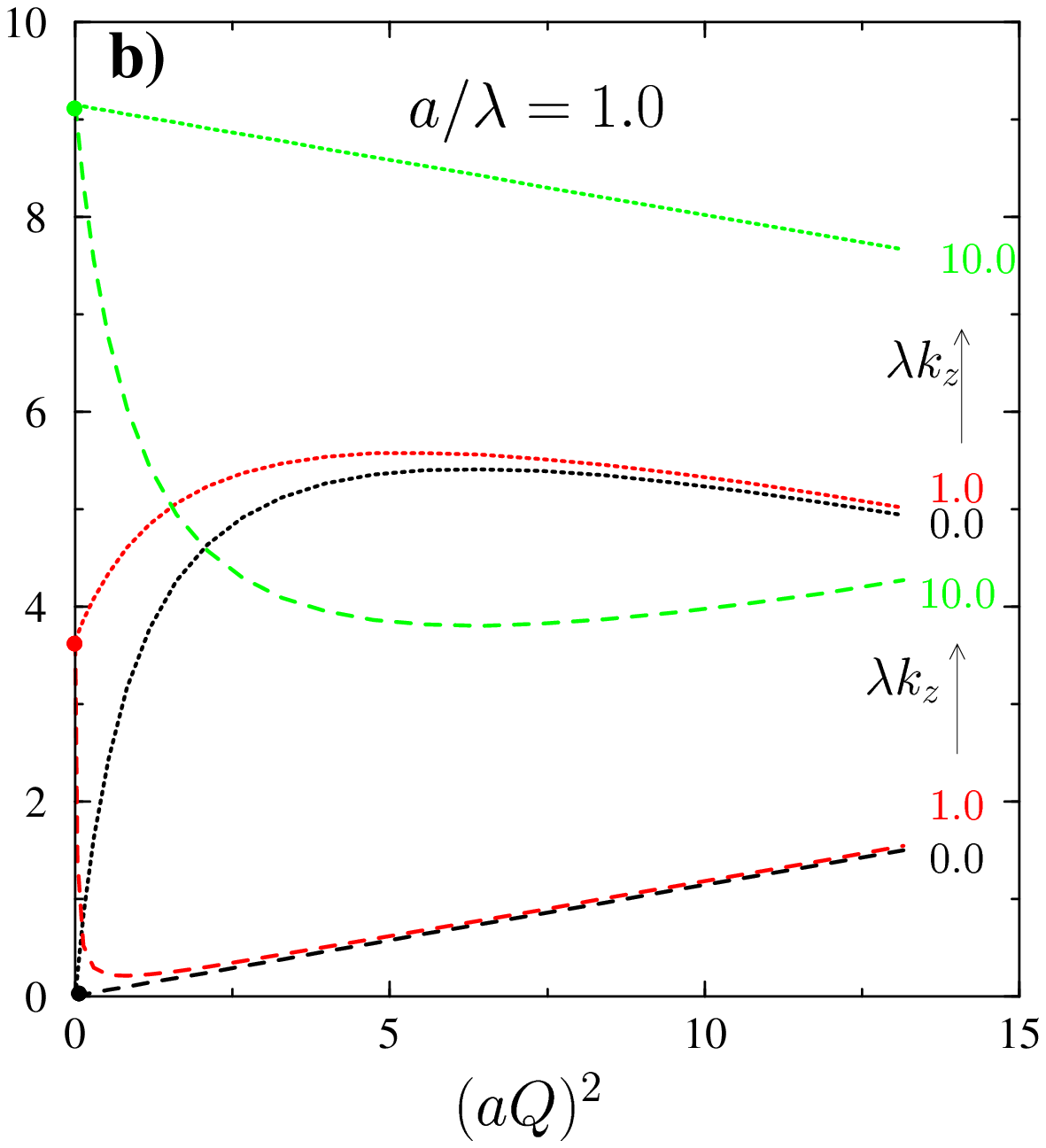}\
\epsfysize=6truecm\epsfbox{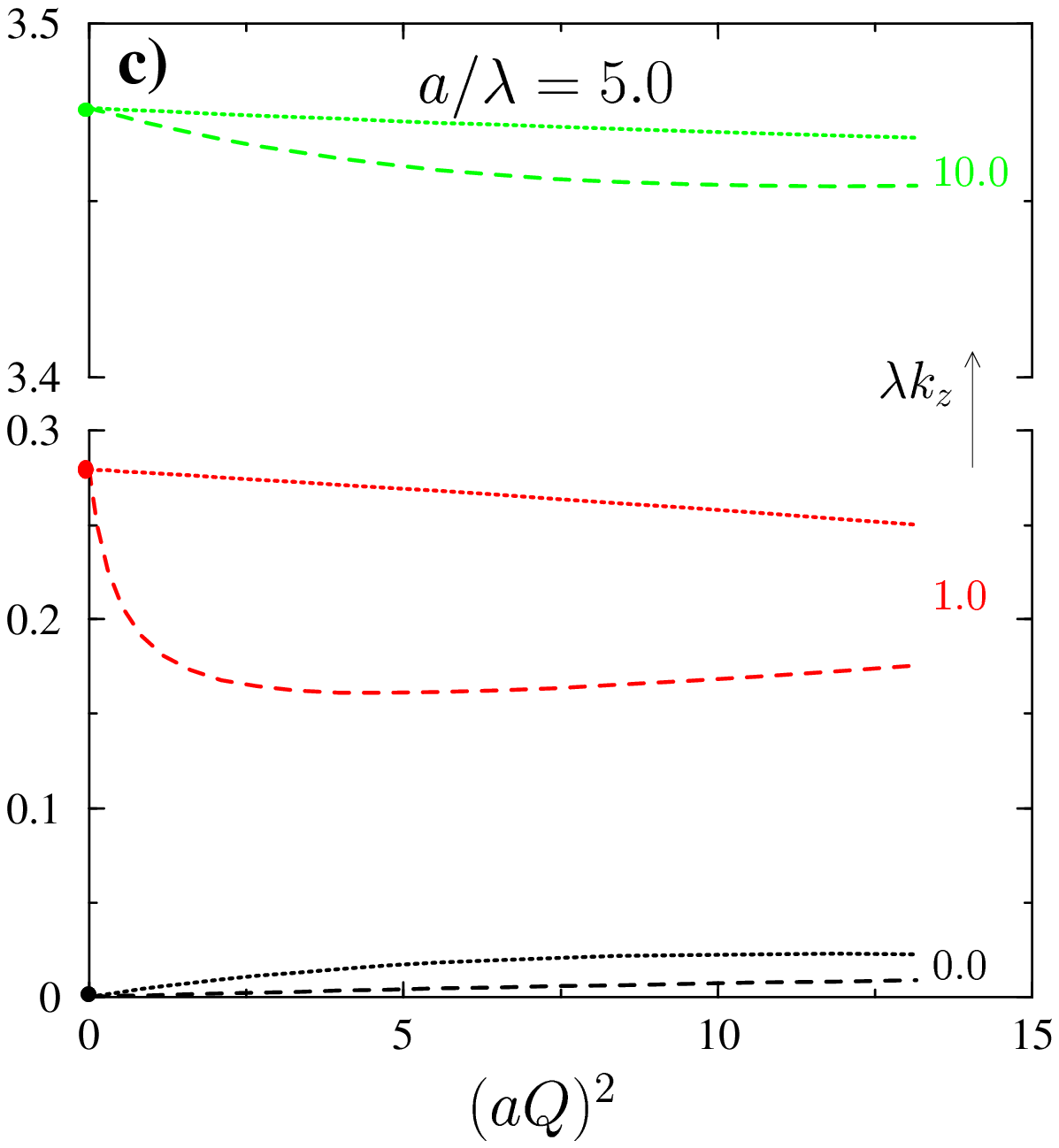}}} 
\medskip
\caption{Elastic eigenvalues as a function of $(aQ)^2$ for
$\alpha=\pi/12$, and a) $a/\lambda=0.2$, b) $a/\lambda=1.0$, and c)
$a/\lambda=5.0$.  The different lines correspond to values of $\lambda
k_z=0.0,1.0,10.0$, and $50.0$ only in plot a).  Dotted and dashed
lines correspond to: longitudinal and transversal eigenvalues
respectively.  \label{cskz} }
\end{figure} 


\begin{multicols}{2}\narrowtext

\begin{figure}
\centerline{\hbox{\epsfysize=6truecm\epsfbox{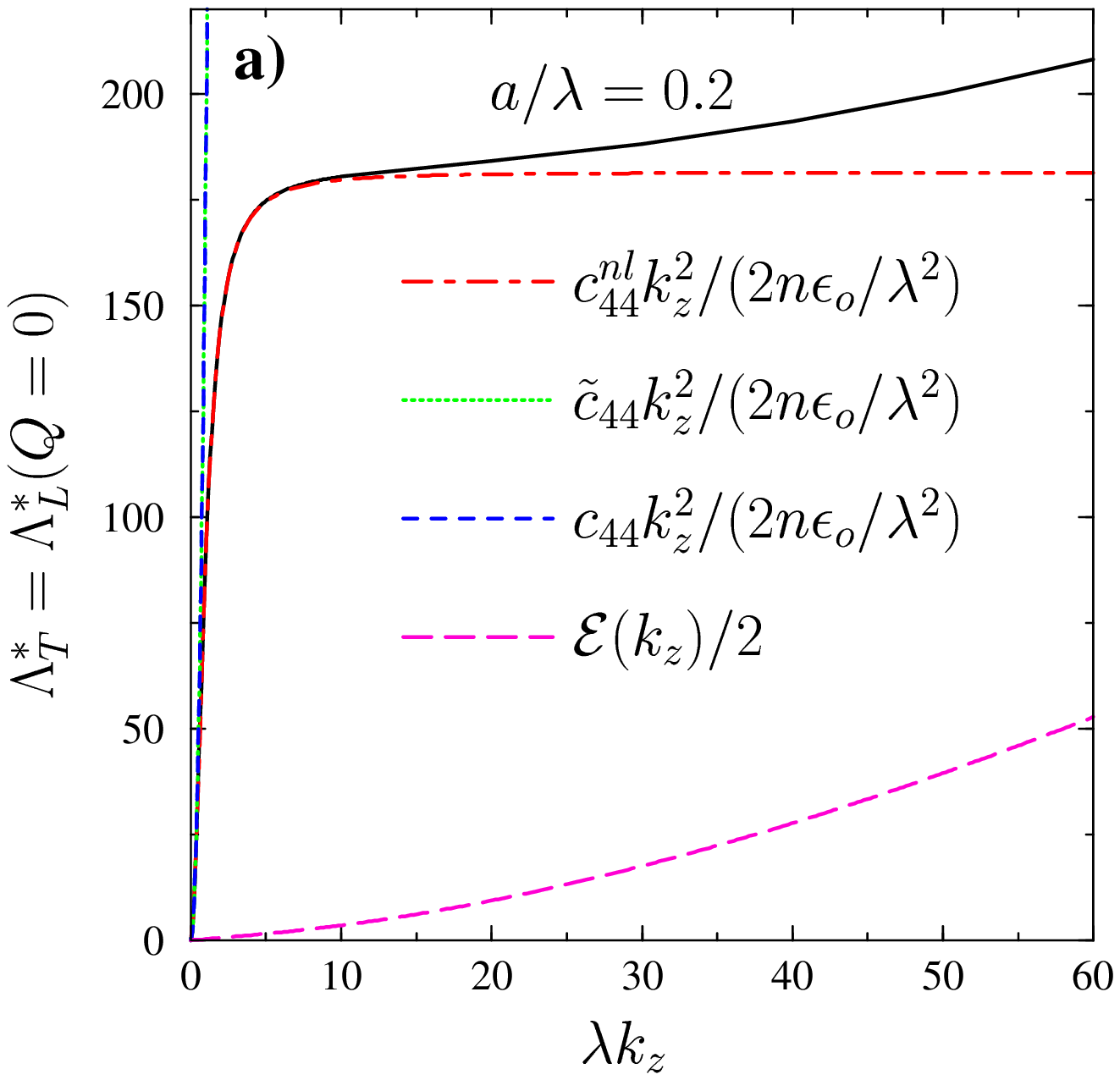}\
\epsfysize=6truecm\epsfbox{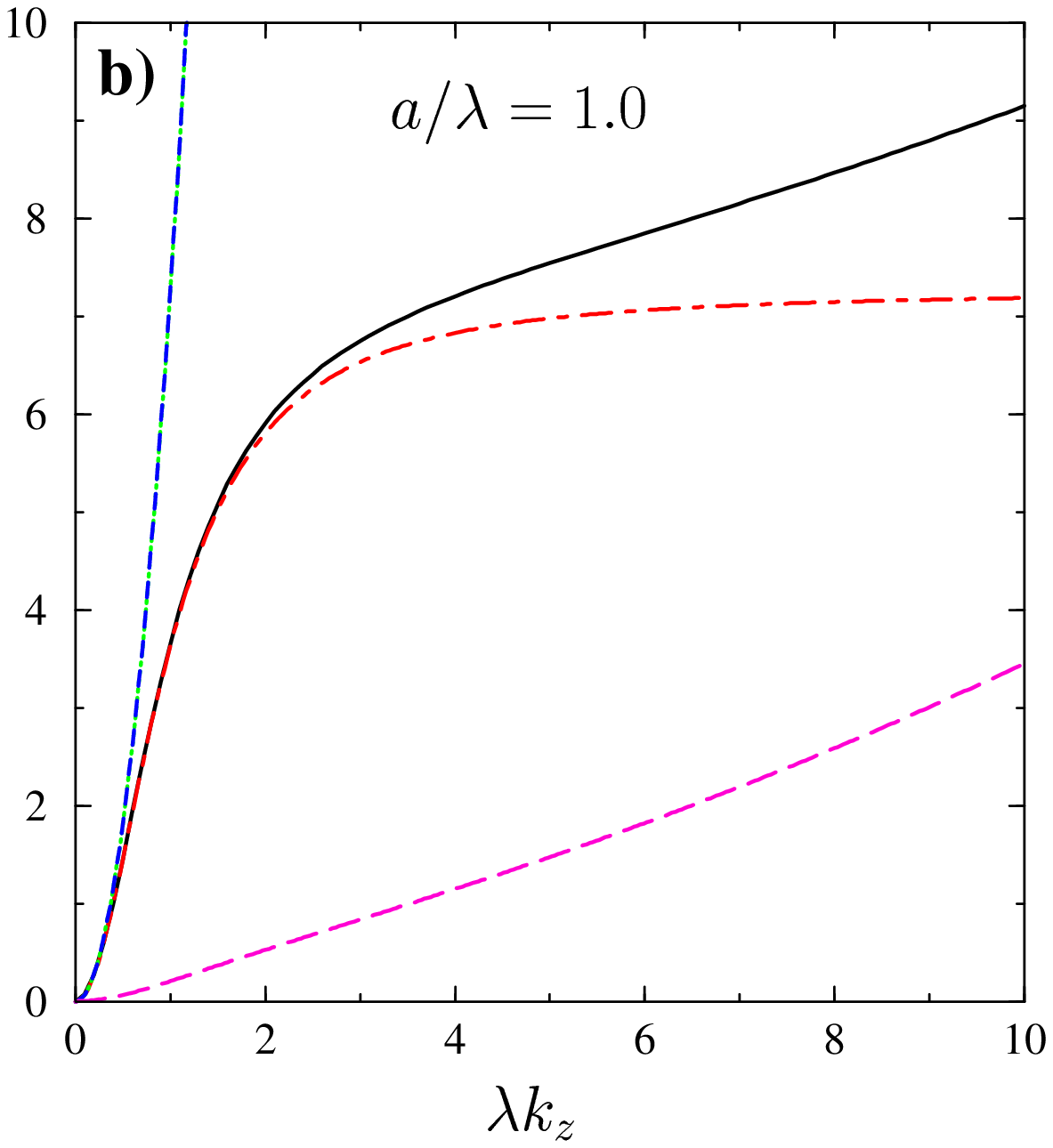}\
\epsfysize=6truecm\epsfbox{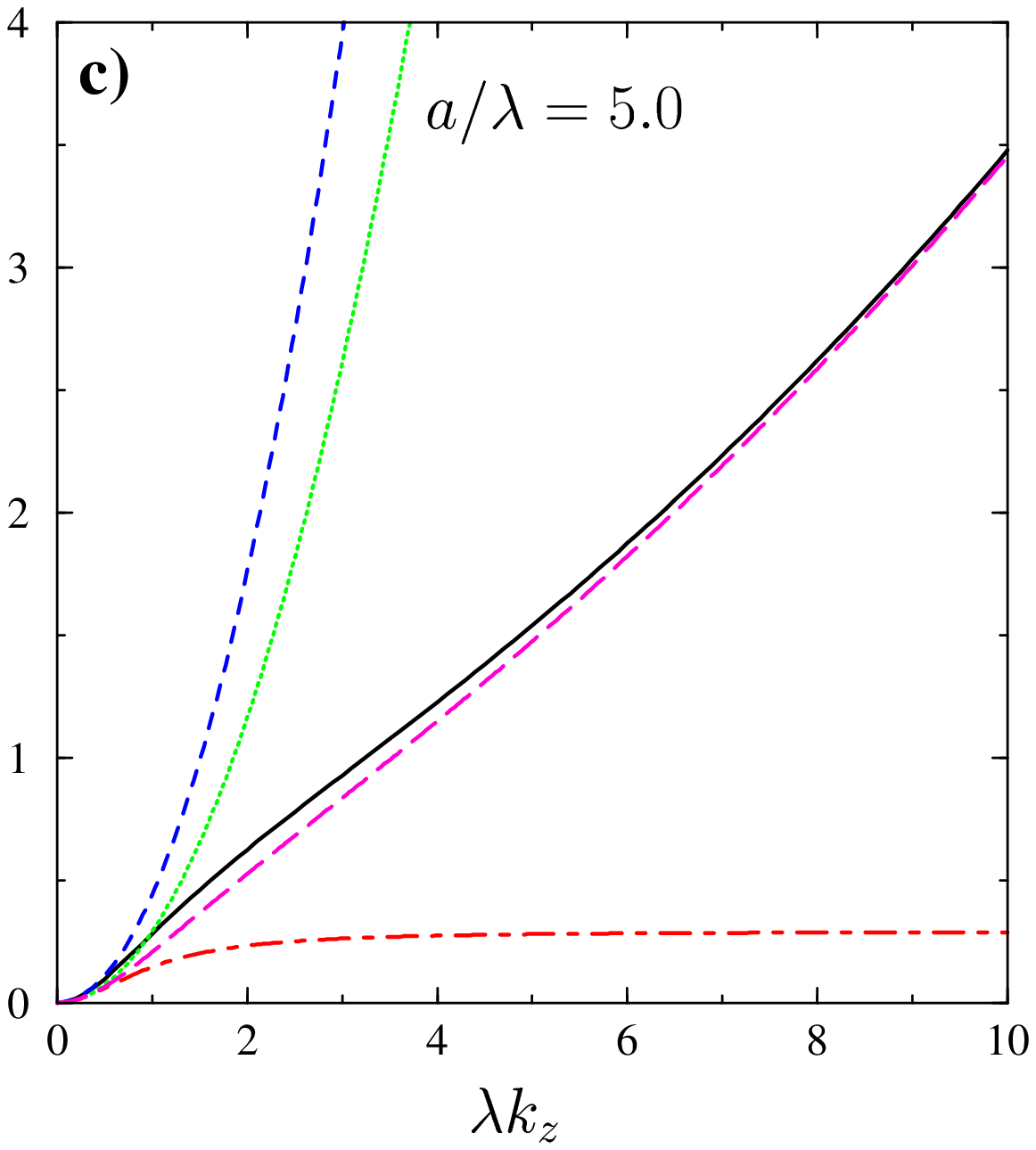}}}
\medskip
\caption{ Elastic eigenvalues, $\Lambda_L(0,k_z)=\Lambda_T(0,k_z)$, as
a function of $\lambda k_z$ for $Q=0$, and a) $a/\lambda=0.2$, b)
$a/\lambda=1.0$, and c) $a/\lambda=5.0$.  The numerical results
indicated by solid lines are compared to the limiting cases in
Sec.~\ref{sec:limits}: The dot-dashed, dotted, dashed, and
long--dashed lines correspond to the nonlocal continuum, local
continuum, local, and single line limits, respectively.
\label{zaxis} }
\end{figure}

To better emphasize this discussion, consider the elastic eigenvalues
along the $Q=0$-axis, i.e. the point highlighted by a black dot in the
previous figures, as a function of $\lambda k_z$.  The results, for
the same values $a/\lambda=0.2$, $1.0$, and $5.0$, are plotted in
Figs.~\ref{zaxis}a) through c), and compared to the limiting cases
described in the previous section.  In particular, the dot-dashed line
in these figures is the outcome of a nonlocal continuum approximation,
whereas the dotted and the dashed lines correspond to the local
continuum and the local results defined in subsection
\ref{sec:local}. The long-dashed line is the graphical representation
of Eq.~(\ref{eq:m4}), i.e., of the single line behavior expected at
very low densities of flux lines.

\end{multicols}\widetext

\begin{multicols}{2}\narrowtext

\subsection{Comparison to analytic approximations}
\label{sec:comparison}

We observe in Fig.~\ref{zaxis} that
$\Lambda_L^*(Q=0,k_z)=\Lambda_T^*(Q=0,k_z)$ increase monotonously with
$\lambda k_z$, with the single-line behavior of ${\cal E}(k_z)$ in
\equ{eq:m4} eventually taking over for values of $\lambda k_z$ greater
than a characteristic $\lambda k_z^c$, proportional to $(\epsilon
a/\lambda)^{-1}$.  The contribution of the nonzero reciprocal lattice
vectors in \equ{eq:m8} becomes relevant for $\lambda k_z>\lambda
k_z^c$.  In the nonlocal continuum approximation, i.e., considering
only $\bbox{G}=\bbox{0}$, $\Lambda_L^*(Q=0,k_z)=\Lambda_T^*(Q=0,k_z)$
reaches a saturation value of $2\pi n^2 \epsilon_o$ at large values of
$\lambda k_z$. Indeed, for high areal densities of $a/\lambda=0.2$,
our numerical outcome is very close to the nonlocal continuum
approximation only up to $\lambda k_z^c$.  On the other hand, at low
areal densities, such as for $a/\lambda=5.0$, the numerical data are
obviously much closer to the single-line limit, than to any of the
other forms, over the whole range of values of $\lambda k_z$.  As
shown in Figs.~\ref{angles}c) and \ref{cskz}c), the variations of
$\Lambda_L$ and $\Lambda_T$ with $(aQ)^2$ in the latter case are very
weak.  The eigenvalues throughout the IBZ essentially coincide with
their values on the $Q=0$-axis, which is set by the form of ${\cal
E}(k_z)$.  In the interval between these two values of $a/\lambda$,
$\Lambda_L(Q=0,k_z)=\Lambda_T(Q=0,k_z)$ gradually cross over from
saturation type behavior to the parabolic-log dependence expressed in
Eq.~(\ref{eq:m4}).  The results of the local approximation naturally
fit well the behavior of both eigenvalues for small values of $Q$ and
$k_z$, as expected.

In discussing Fig.~\ref{cskz}, we noted that while the non-local
continuum expressions capture the qualitative form of the numerical
results, there are also important quantitative differences.  Using the
insights gained from the numerics, we shall now present an analytic
form that corrects some of these discrepancies.  One difference from
the continuum results arises from the limiting slope in
Fig.~\ref{cskz}: In the continuum limit, the term
$\tilde{c}_{66}Q^2/2$ eventually determines the slope of the linear
regime of the transversal eigenvalues.  However, as shown in
Fig.~\ref{compressheartilt}a), the actual shear modulus $c_{66}$ can
be quite different from the limiting value of $\tilde{c}_{66}$ used in
the continuum approximation.  The differences in slope thus merely
reflect the differences between $c_{66}$ and $\tilde{c}_{66}$ as a
function of $a/\lambda$, as already discussed in Sec.~\ref{sec:local}.
This deficiency of the continuum eigenvalues is thus removed by using
the exact value of $c_{66}$ from Fig.~\ref{compressheartilt}a).

A second difference is an overall shift of the eigenvalues from the
continuum limit prediction, which becomes more pronounced at larger
$\lambda k_z$.  This clearly originates in the differences appearing
already in Figs.~\ref{zaxis}a) and b) for
$\Lambda_L(Q=0,k_z)=\Lambda_T(Q=0,k_z)$ at $Q=0$ and sufficiently
large $k_z$.  Since the continuum approximation uses only the term
with $\bbox{G}= 0$ in Eq.~(\ref{eq:m8}), we introduce a correction by
replacing the sum over all the remaining reciprocal lattice vectors
with an integral.  These corrections result in the following
expression for the transversal eigenvalue:
\begin{eqnarray}
  \label{eq:n3} & & \Lambda_T(Q,k_z)\simeq \frac{1}{2}[c_{66}
  Q^2 + c_{44}^{nl} k_z^2] + \frac{B^2}{8\pi} (\delta_{\alpha
  \beta}-\hat{Q}_{\alpha}\hat{Q}_{\beta}) \nonumber \\ & &\times
  \int^{\ '} \frac{d^2 \bbox{G}}{{\cal A}_{BZ}} [G_{\alpha
  \beta}(\bbox{G},k_z)- G_{\alpha \beta}(\bbox{G},0)].
\end{eqnarray}
In this equation, the correct local shear modulus, which depends on
the field strength or lattice spacing $a$ through Eq.(\ref{eq:l6}), is
used in place of its continuum limit.  Also the non-linear modulus
$c_{44}^{nl}$ is corrected by the additional integral over the nonzero
reciprocal lattice vectors, properly normalized by the BZ area
$A_{BZ}=8\pi^2/(\sqrt{3} a^2)$.  To exclude the point at $\bfm G=0$,
the radial component of $\bbox{G}$ in the above integral goes from
$C$, defined by $A_{BZ}=\pi C^2$, to $\infty$ (or, if necessary, to
the short distance cutoff $\xi^{-1}$).  After evaluating the integral,
we obtain
\begin{eqnarray}
  \label{eq:n4} 
  & & \Lambda_T(Q,k_z)\simeq \frac{1}{2}[c_{66}
  Q^2 + c_{44}^{nl} k_z^2] \nonumber \\ & &
  +\frac{n\epsilon_o}{4}\epsilon^2 k_z^2
  \ln\left(\frac{\kappa^2/\epsilon^2+ \lambda^2 k_z^2+1}{\lambda^2
  C^2/\epsilon^2+ \lambda^2 k_z^2+1}\right) \nonumber \\ & &
  +\frac{n\epsilon_o}{4\lambda^2}\ln\left(\frac{\lambda^2 k_z^2 +
  \lambda^2 C^2+1}{\lambda^2 C^2+1}\right).
\end{eqnarray}
We have ascertained that this expression provides an excellent fit to
the numerically obtained results for $\Lambda_T(Q,k_z)$ along the
$\alpha=\pi/12$ direction.  In the next section we provide explicit
comparisons of the analytic expression and numerics for different
values of the anisotropy.  The longitudinal eigenvalue along the
$\alpha=\pi/12$ direction can also be reasonably well fitted by a
similar analytic expression, which replaces the term
$[c_{66}Q^2+c_{44}^{nl} k_z^2]/2$ in Eq.~(\ref{eq:n4}) with the
expression in \equ{eq:n2}, with a further substitution of
$c_{66}Q^2/2$ for $n\epsilon_oQ^2/8=\tilde{c}_{66}Q^2/2$, as
\begin{eqnarray}
  \label{eq:n5} 
  & & \Lambda_L(Q,k_z)\simeq 2\pi n^2 \epsilon_o
  \frac{\lambda^2(Q^2+k_z^2)}{[\lambda^2(Q^2+k_z^2)+1]}
  -\frac{1}{2} c_{66}Q^2 \nonumber \\ & &
  +\frac{n\epsilon_o}{4}\epsilon^2 k_z^2
  \ln\left(\frac{\kappa^2/\epsilon^2+ \lambda^2 k_z^2+1}{\lambda^2
  C^2/\epsilon^2+ \lambda^2 k_z^2+1}\right) \nonumber \\ & &
  +\frac{n\epsilon_o}{4\lambda^2}\ln\left(\frac{\lambda^2 k_z^2 +
  \lambda^2 C^2+1}{\lambda^2 C^2+1}\right).
\end{eqnarray}
\subsection{Anisotropy}
\label{sec:anisotropy}

We conclude this section by discussing the dependence of our results
on the anisotropy of the superconductor.  The results presented so far
correspond to a particular value of the anisotropy parameter, namely
$\epsilon=0.1$, which falls within the range (from $1/10-1/5$)
reported for ${\rm YBa_2Cu_3O_7}$ (YBCO) in the literature
\cite{Blatter94}.  However, smaller values of $\epsilon$, in the
interval $1/100-1/50$, characterize highly anisotropic materials such
as ${\rm Bi_2Sr_2CaCu_2O_8}$ (BiSCCO).  For such highly anisotropic
materials, the discreteness of their layered structure becomes
important, and one may well question the validity of a three
dimensional Landau--Ginzburg description.  An alternative model
is a set of weakly (Josephson) coupled superconducting layers.  The
applicability of the three dimensional description is typically
assessed by comparing the coherence length along the $c$ axis
$\xi_c=\epsilon \xi$, with the distance $d$ between the Cu-O layers in
the material\cite{Blatter94}.  A coherence length $\xi_c$ larger than
the layer spacing usually justifies a continuous description along the
$c$ direction \cite{ccutoff}.  This is certainly the case for YBCO,
but not necessarily for BiSCCO.  Nevertheless, extrapolating the
results of the continuous description may also provide insights into
the elastic properties of highly anisotropic superconductors such as
BiSCCO.  To this end, we compare results obtained for three different
values of the anisotropy parameter, namely $\epsilon=0.02,0.1$, and $0.5$,
for the usual areal densities of $a/\lambda=0.2,1.0,5.0$.

\begin{figure}
\centerline{\hbox{\epsfysize=6truecm\epsfbox{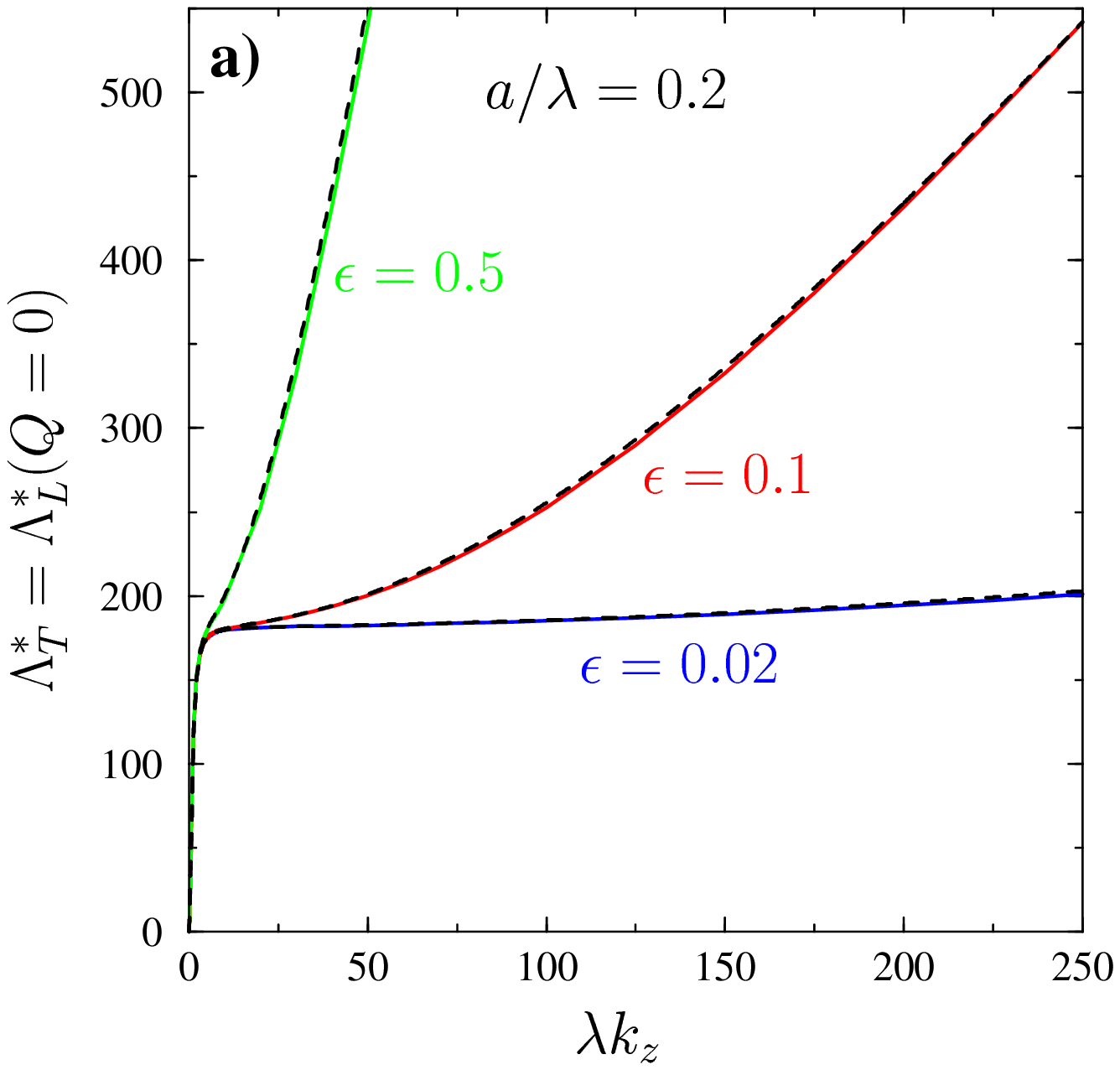}\
\epsfysize=6truecm\epsfbox{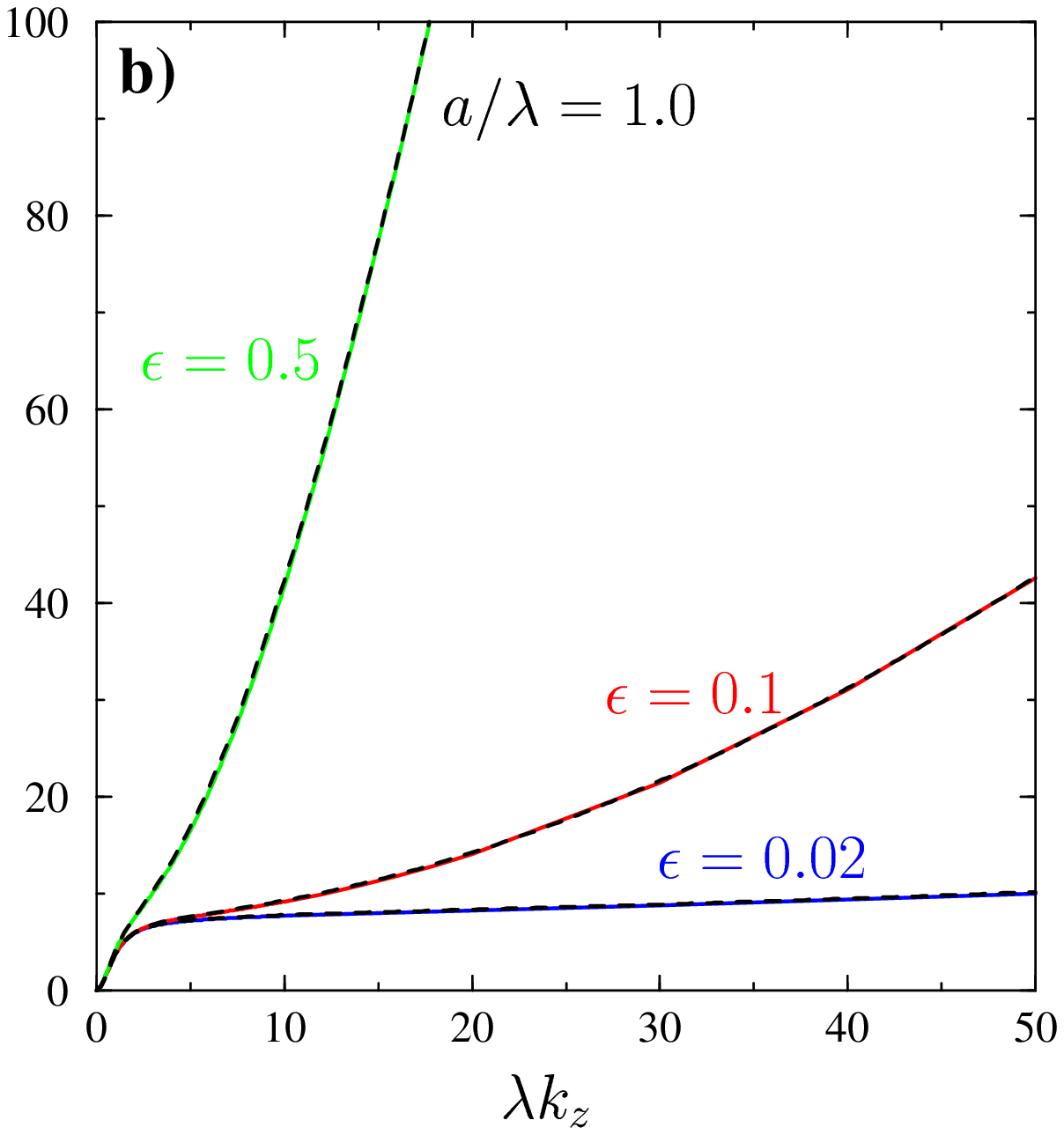}\
\epsfysize=6truecm\epsfbox{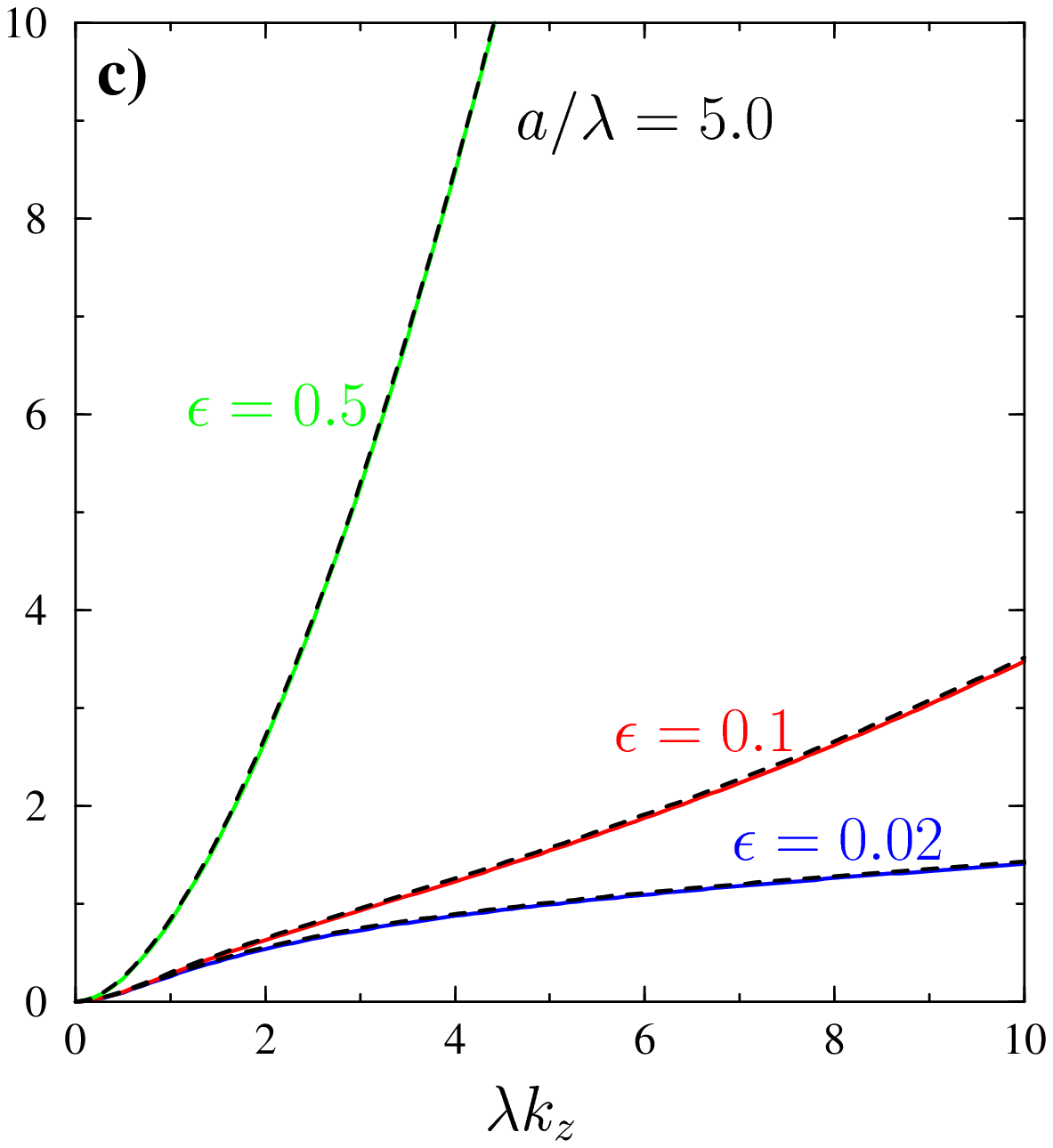}}} 
\medskip
\caption{ Elastic eigenvalues along the $Q=0$-axis as a function of
$\lambda k_z$, with different values, $\epsilon=0.02$, $0.1$, and
$0.5$, of the anisotropy parameter. The dashed lines depict the
analytic expression in \protect{\equ{eq:n4}} with the corresponding
choice of parameters.\label{epsilons}}
\end{figure}

\begin{figure}
\centerline{\hbox{\epsfysize=6truecm\epsfbox{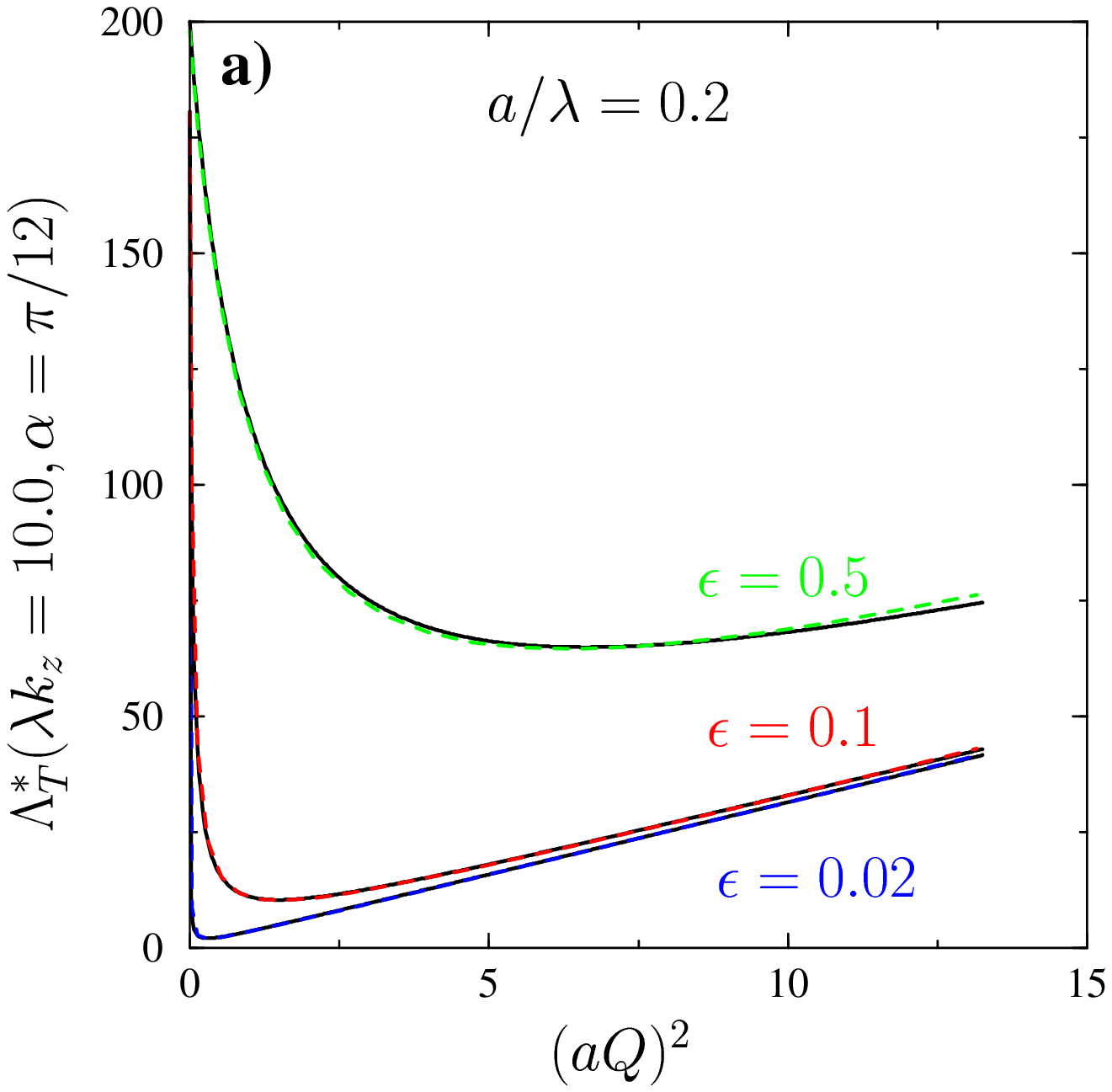}\
\epsfysize=6truecm\epsfbox{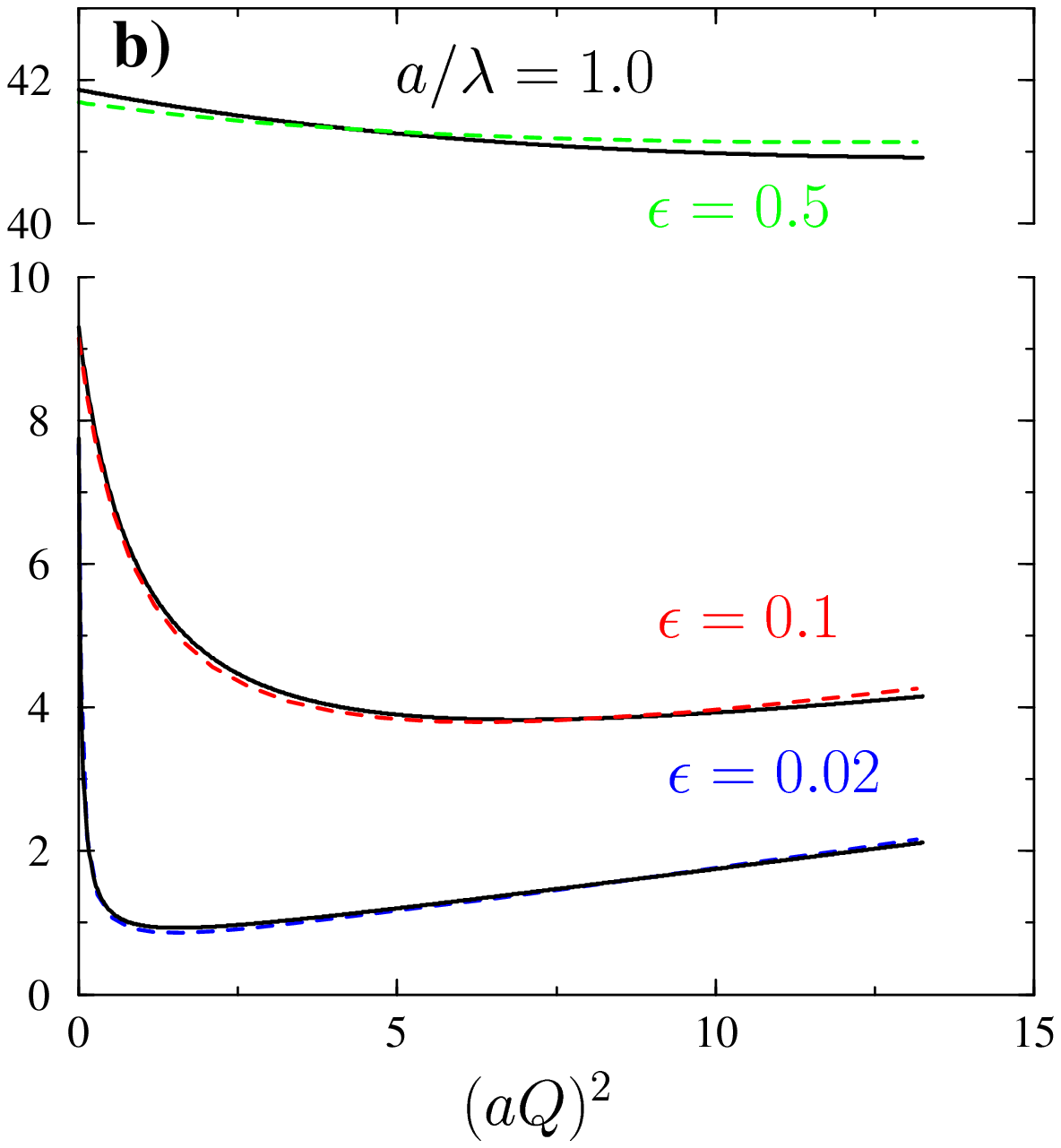}\
\epsfysize=6truecm\epsfbox{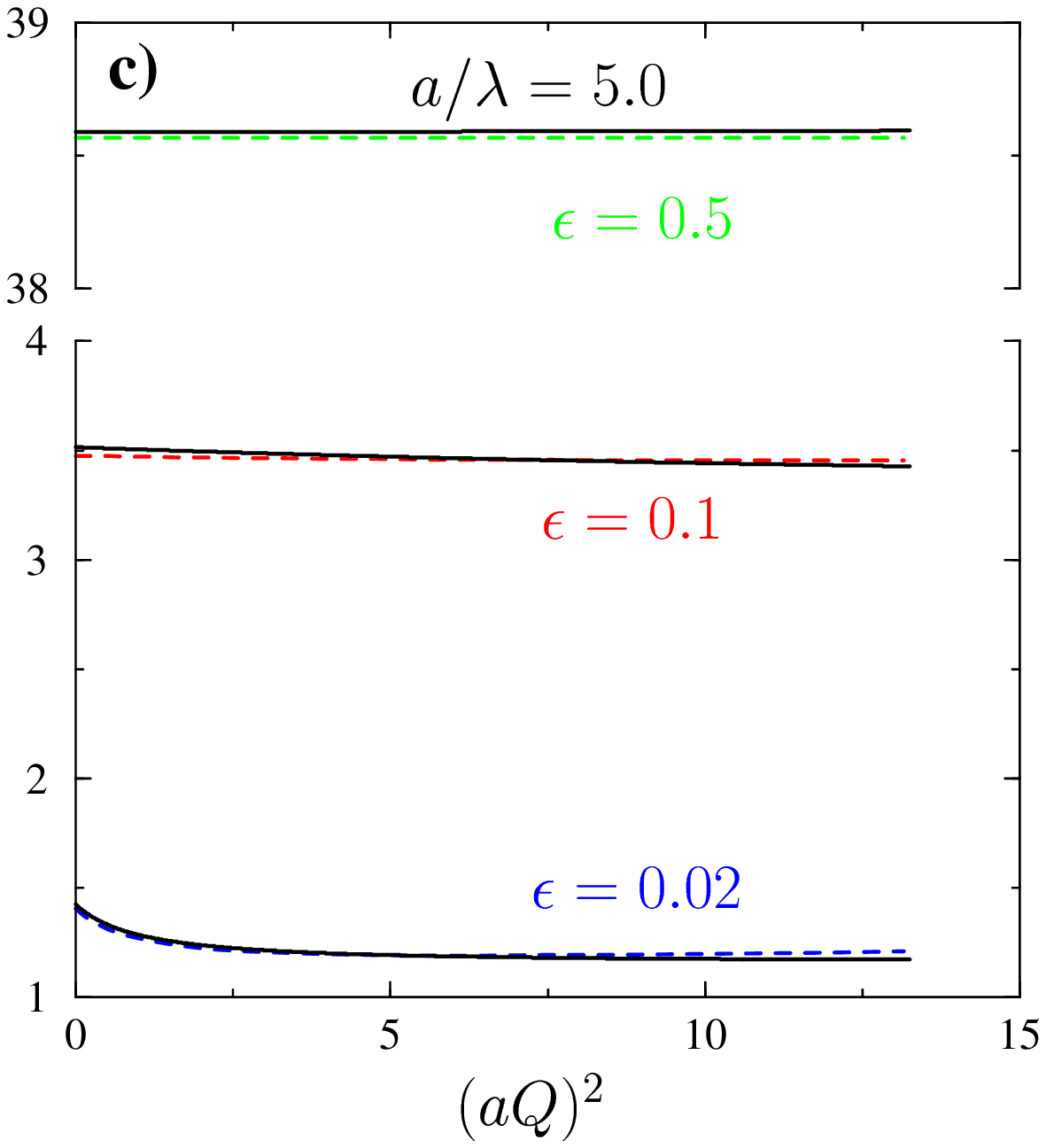}}} 
\medskip
\caption{ Transversal eigenvalue along the $\alpha=\pi/12$ direction
in the IBZ as a function of $(aQ)^2$, and for $\lambda k_z=10.0$.  The
dashed lines are numerical results for $\epsilon=0.02$ (BiSCCO), $0.1$
(YBCO), and $0.5$, while the solid lines depict the corresponding
results from the analytic expression in
\protect{\equ{eq:n4}}. \label{sepsilons}}
\end{figure}

In Figs.~\ref{epsilons} and \ref{sepsilons}, we again plot the elastic
eigenvalues along the $Q=0$-axis, and the transversal eigenvalue along
the $\alpha=\pi/12$ direction inside the IBZ, respectively, for
different choices of $\epsilon$.  There is clearly a strong dependence
on $\epsilon$, mainly originating from the $\epsilon$-dependence of
the line tension ${\cal E}(k_z)$ in \equ{eq:m4}.  Naturally, the most
pronounced tendency is the decrease in eigenvalues with $\epsilon$,
reflecting the general softening of the vortex lattice. It is worth
pointing out however, that the slopes of the linear part of the curves
in Fig.~\ref{sepsilons} are independent of $\epsilon$, reflecting
merely the fact that the shear modulus $c_{66}$ is independent of the
anisotropy parameter. The dashed lines in the three graphs of
Fig.~\ref{epsilons} correspond to the analytic expression of
\equ{eq:n4} (for the appropriate parameters $a/\lambda$ and
$\epsilon$), and clearly provide an excellent fit to the numerical
data along the $Q=0$ axis.  Similarly in Fig.~\ref{sepsilons}, the
solid lines are from \equ{eq:n4} for the transveral eigenvalue, again
showing fairly good agreement with the numerical data.  Analogous
results are obtained for the longitudinal eigenvalue (not shown).

\end{multicols}\widetext

\begin{multicols}{2}  \narrowtext

\section{The Melting line}
\label{sec:melting}

The melting of a vortex lattice by thermal fluctuations has attracted
considerable attention in the context of high temperature
superconductors.  This transition has been observed by means of
several experimental techniques such as bulk magnetization, local
induction, and latent heat measurements
\cite{Gammel88,Safar92,Kwok92,Pastoriza95,Zeldov95,Liang96,Welp96,Roulin96,Oral98}.
The {\it line-like} nature of the constituent elements provides
intriguing challenges to theoretical analysis.  Important features of
the melting transition are the negative slope of the melting curve
$T_m(B)$ at high fields, its reentrant behavior at low fields, and its
marked dependence on anisotropy.  Some of these features can be
extracted from simple models of the vortex lattice, as in the
so-called {\it XY} \cite{TH}, {\it Bose} \cite{RMN}, and {\it cage}
\cite{Nelson94} models.  These models agree in the prediction of
certain universal features, such as the scaling of the melting
temperature with the anisotropy parameter, or the magnetic field, in
the high field region of the phase diagram\cite{Koshelev98}.

In the absence of a rigurous theory for three-dimensional melting, the
position and shape of the vortex lattice melting line is usually {\it
estimated} using the so-called Lindeman criterion.  According to this
criterion, the lattice melts when thermally induced fluctuations of a
lattice point become comparable to the lattice spacing.  This
condition can be written as
\begin{equation}
  \label{eq:me1}
  [\langle u^2(r_0)\rangle]^{1/2} \sim c_L a,
\end{equation}
where $c_L\sim 0.1-0.2$ is the (empirically chosen) Lindeman parameter.
The extent of fluctuations is measured by the autocorrelation function,  
$\langle u^2(r_0)\rangle$. This quantity was calculated by Nelson and Seung
\cite{Nelson89} in the local limit, and by Brandt \cite{Brandt89} and
Houghton~{\it et al.}  \cite{Houghton89} for the nonlocal continuum
limit, and  in more general circumstances including variations of the
amplitude of the order parameter in the Ginzburg-Landau hamiltonian.

In the preceeding sections, we calculated the harmonic energy cost of
fluctuations of the vortex lattice.  In a classical equilibrium state,
each independent harmonic mode acquires a thermal energy of $k_BT/2$.
By adding the corresponding squared amplitudes of the normal modes,
the autocorrelation function is obtained as
\begin{equation}
  \label{eq:me2}
  \langle |{\bf u}({\bf r}_0)|^2 \rangle=\frac{k_BT}{2}\int
\frac{dk_z}{(2\pi)}\int_{BZ}\frac{d^{2}{\bf Q}}{(2\pi)^2}\
(\Lambda_L^{-1} +\Lambda_T^{-1}).
\end{equation}
The Lindeman criterion now provides a rough estimate of the melting
temperature $T_m$, through the relationship
\begin{equation}
  \label{eq:me3}
 \frac{k_BT_m}{2} \int
\frac{dk_z}{(2\pi)}\int_{BZ}\frac{d^{2}{\bf Q}}{(2\pi)^2}\
(\Lambda_L^{-1} +\Lambda_T^{-1}) \sim c_L^2 a^2.
\end{equation}

The leading contribution to the fluctuations comes from the
transversal modes, which as discussed in Sec.~\ref{sec:limits} are
always smaller than the longitudinal modes.  We also ignore the
angular dependence observed in Figs.~\ref{angles}, which is rather
weak, and should not give rise to qualitatively different results.
Furthermore, we assume that the transversal eigenvalue is isotropic,
and given by \equ{eq:n4}.  This expression can be integrated
analytically over the two-dimensional BZ, resulting in a rather long
expression, that we have finally integrated numerically over $\lambda
k_z$. The numerical integral over $\lambda k_z$ is, in all cases,
between a long wavelength cutoff equal to $10^{-6}$ and the
$\epsilon$-dependent short scale cutoff of
$\lambda/\xi_c=\kappa/\epsilon$ \cite{ccutoff}.

Provided that we are sufficiently far away from the critical
temperature $T_c$, one can further assume that the penetration length
is independent of temperature, i.e. $\lambda(T) \sim \lambda(0)$. As
we get closer to $T_c$, however, one must consider the temperature
dependence of $\lambda$ in order to obtain sensible results. For the
sake of consistency of our analysis, within the Ginzburg-Landau
critical regime, we then use the mean-field temperature dependences of
both $\lambda$ and $\xi$, i.e. $\lambda(T)=\lambda(0)(1-T/T_c)^{-1/2}$
and $\xi(T)=\xi(0)(1-T/T_c)^{-1/2}$. We expect this approximation to
fail in close vicinity of $T_c$ due to critical fluctuations.

\renewcommand{\caption}[1]{\refstepcounter{figure}\protect\noindent%
  \protect\parbox{8.6cm}{\small FIG. \thefigure. #1}}

\begin{figure}
\centerline{\epsfxsize=8cm\epsfbox{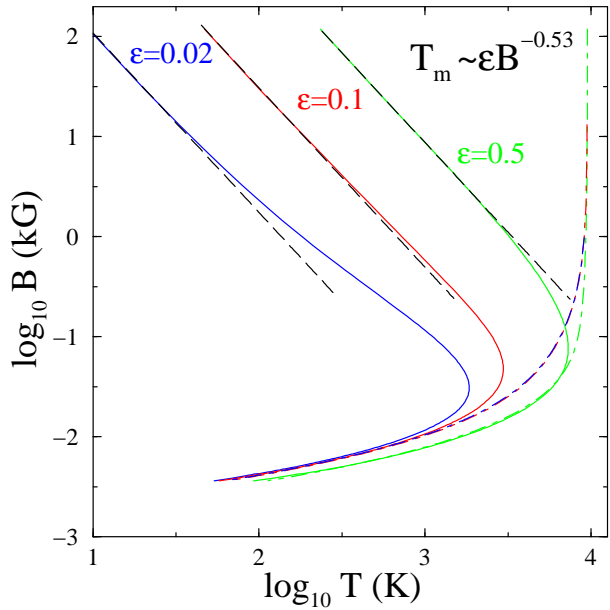}}
\caption{The melting line $T_m(B)$, for $\epsilon=0.02,0.1,0.5$, from
the Lindeman criterion, assuming {\em a temperature independent}
penetration depth $\lambda$. At high fields, the melting temperature
decreases with the magnetic induction as $B^{-0.53}$, as indicated by
the asymptotic dashed lines, and linearly with the anisotropy
parameter $\epsilon$. The dotted-dashed lines show the melting lines
resulting from the local elastic moduli.  While the latter agrees
quite well with the reentrant low field part of the curve, it yields a
very different behavior at higher fields.\label{melting0}}
\end{figure}

\begin{figure}
\centerline{\epsfxsize=8truecm\epsfbox{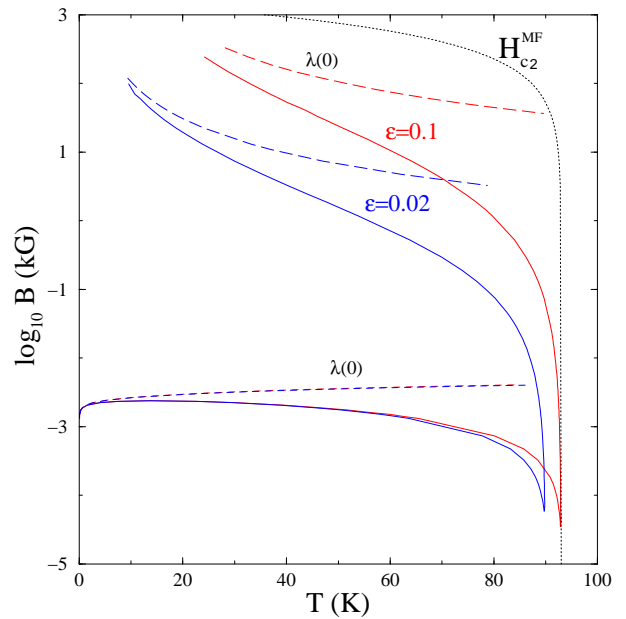}}
\caption{The melting line $T_m(B)$, for $\epsilon=0.02,0.1$, assuming
{\em a mean-field temperature dependent} penetration depth $\lambda$.
The dashed lines show the results obtained assuming
$\lambda=\lambda(0)$ up to the point in which they are still
reasonable. \label{melting1}}
\end{figure}
\begin{figure}
\centerline{\epsfxsize=8truecm\epsfbox{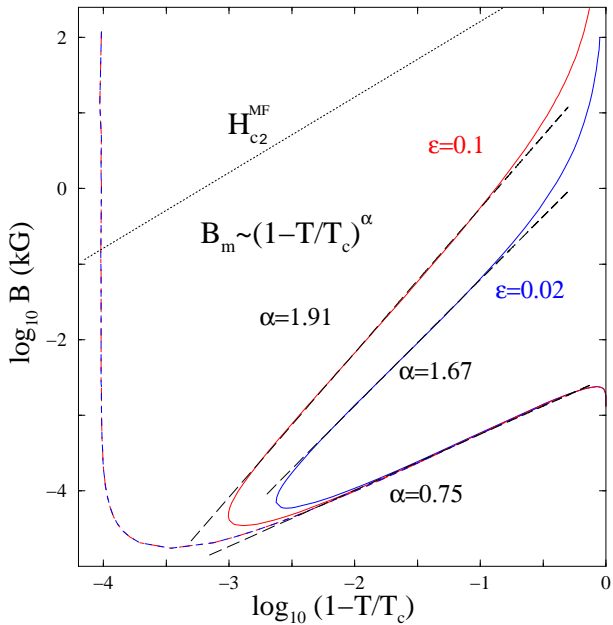}}
\caption{The low-field region of the reentrant melting line $B_m$,
plotted as a function of $(1-T/T_c)$, for $\epsilon=0.02,0.1$.  Close
to the critical temperature $T_c$, the melting induction grows as a
power of the reduced temperature $1-T/T_c$, as indicated by the
asymptotic dashed lines.  As in Fig.~\ref{melting0}, the dotted-dashed
lines show the melting lines resulting from the local elastic
moduli.\label{melting2}}
\end{figure}

In Figs.~\ref{melting0}, \ref{melting1}, and \ref{melting2}, we depict
the resulting melting lines for different values of the anisotropy
parameter $\epsilon=0.02$ (BiSCCO), and $\epsilon=0.1$ (YBCO). Note
that some of the data are presented in a log--log or in a log--linear
plot in order to emphasize the scaling with the magnetic field in the
high field region, as well as to better visualize the lower part of
the melting line, i.e. the reentrant portion of the phase diagram. The
dotted line in the last two figures represents the upper critical
field $H_{c_2}^{MF}(T)=\phi_o(1-T/T_c)/2\pi\xi^2(0)$, with
$T_c=93^\circ K$ and $\xi(0)=14 \AA$. Close to $H_{c_2}^{MF}(T)$ the
amplitude of the order parameter is strongly reduced due to the
overlap of the vortex cores and these results are no longer
valid. However, well below this line, the London limit provides a good
description of the system.  In Fig.~\ref{melting0}, we represent the
results obtained assuming $\lambda(T)=\lambda(0)=1400 \AA$, for
$\epsilon=0.02,0.1,0.5$. At high fields $a/\lambda\ll1$, the melting
temperature decreases as $B^{-0.53}$ in all cases, consistent with the
prediction of $T_m \sim B^{-1/2}$, common to the above mentioned
models \cite{TH,RMN,Nelson94}.  We have also checked that the melting
temperature decreases linearly with $\epsilon$, for various values of
$a/\lambda$ within the high field region, i.e. $T_m\sim \epsilon
B^{-0.53}$. This is another feature which is in agreement with
previous predictions, and with available experimental data.  In this
figure we also encounter some unreasonably high melting temperatures
which are the consequence of the assumption $\lambda(T)=\lambda(0)$.

Figure \ref{melting1} shows the results obtained for the melting line
$T_m$ assuming $\lambda(T)=\lambda(0)(1-T/T_c)^{-1/2}$. We have used
the values of $T_c=90$ for BiSCCO ($\epsilon=0.02$), and $T_c=93$ for
YBCO ($\epsilon=0.1$). As in the previous figure, within the high
field region of the phase diagram, the melting temperature decreases
as the external field is increased. Quantitative and qualitative
differences between the new curves and the corresponding results for
$\lambda(T)=\lambda(0)$ (depicted as dashed lines in the figure) can
be observed even at low temperatures.  Near the critical temperature,
the low-field portion of the melting curves has a negative slope (as
for their high field counterparts), but at low temperatures the slope
is positive, so that the melting temperature decreases with the
external field at very low densities of flux lines.

At the scale of Fig.~\ref{melting1}, the high and low field portions
of the melting curve $T_m(B)$ appear almost discontinuous close to
$T_c$.  However, as indicated in the blow-up of this region in
Fig.~\ref{melting2}, this is in reality a sharp but continuous turn
around.  Following the literature \cite{Liang96,Welp96,Roulin96}, we
fit the melting fields $B_m(T)$ to power laws in the reduced
temperature $t\equiv1-T/T_c$, with an exponent of $\alpha$.  The
dashed lines in Figure \ref{melting2} indicate the extent of this
power law regime for each value of $\epsilon$.  We note that even
though we are using the input value of $T_c$, the power law regime
extends at most over two decades.  We naturally obtain different
exponents for the upper and lower branches of the melting curve.  The
apparent exponent of the upper branch actually depends on the
parameter $\epsilon$, taking the value of
$\alpha_{+}(\epsilon=0.1)=1.91$ for `YBCO', and a slightly smaller
value of $\alpha_{+}(\epsilon=0.02)=1.67$ for `BiSCCO'.  Both curves
overlap in the reentrant region at very dilute concentrations where we
can fit to an effective exponent of $\alpha_{-}=0.75$.

The phase diagram of Fig.~\ref{melting1} is qualitatively similar to
the predictions of Ref.~\onlinecite{Fisher91}, where the power law
forms for the phase boundary close to $T_c$ were also first discussed.
In particular, it is reasonably straightforward to estimate the shape
of the phase boundary in the low field region, where the leading
contributions to the transversal modes $\Lambda_T$ come from the shear
modulus $c_{66}$ (which decays exponentially with $a/\lambda$, and
does not depend on $\epsilon$), and from the last term in
Eq.~(\ref{eq:n4}).  As expected for very low densities of flux lines,
the last two terms in this equation tend to the single line tension
${\cal E}(k_z)$ (see \equ{eq:m4}) which, for small values of $k_z$, is
in turn dominated by the $\epsilon$-independent magnetic contribution.
This leads to $B_m^{-}\sim \lambda(T)^{-2}\ln^{-2}(\lambda(T))$, and
consequently to an exponent of $\alpha_{-}=1$ with logarithmic
corrections \cite{Fisher91}.  The effective exponent of
$\alpha_{-}\approx 0.75$ obtained in Fig.~\ref{melting2} is different
from the $\alpha_{-}=1$ expected on the basis of our mean-field form
for $\lambda(T)$ indicating the difficulty of determining the true
exponent from such fits.

For the upper portion of the mean-field line, the melting field is
estimated \cite{Fisher91} to scale as $B_m \sim \lambda(T)^{-4}$,
leading to $\alpha_{+}=2$ for our choice of a mean-field $\lambda(T)$.
While the observed value of $\alpha_{+}(\epsilon=0.1)=1.91$ is not far
from this expectation, the smaller value of
$\alpha_{+}(\epsilon=0.02)=1.67$ indicates the difficulty of finding a
good asymptotic regime.  Naturally, in close vicinity of the critical
point we can no longer use the mean field forms for the divergence of
$\lambda\propto\xi\propto |t|^{-1/2}$.  Using the scaling forms,
$\xi\propto |t|^{-\nu}\propto \lambda^2$, with the XY critical
exponent of $\nu\approx 2/3$, leads to 
$\alpha_{+}=2\alpha_{-}=2\nu$\cite{Fisher91}. Indeed, the values of
the exponent of the upper branch ($\alpha_{+}\sim 1.33-1.36$) reported
from measurements of the melting transition in single crystals of YBCO
\cite{Liang96,Welp96,Roulin96}, are consistent with this prediction.
However, given the difficulties of extracting exactly known exponents
from the data of Fig.~\ref{melting2}, we may well question the
robustness of this procedure.

We would like to emphasize that most of these results cannot be
obtained from the local values of the elastic moduli given in
Sec.~\ref{sec:local}.  The results obtained from such a local elastic
approximation are plotted as dot--dashed lines in 
Figs.~\ref{melting0} and \ref{melting2}.
The latter curves overlap in the high field region, where neither the
local tilt, nor the shear modulus depend on $\epsilon$.  
Rather than decreasing as $T_m\sim B^{-1/2}$, the resulting melting 
temperature reaches a constant value for high $B$, i.e. $T_m\sim
B^{0}\epsilon^{0}$.  On the other hand, the local melting temperature
provides a very good approximation to the melting line at very dilute
concentrations, in the reentrant low field portion of the curve.

\section{Conclusions}
\label{sec:conclu}

In conclusion, we have obtained the elastic moduli of the flux line
lattice in a systematic and detailed way which is not restricted to
the most commonly considered continuum limit.  The transversal and
longitudinal harmonic eigenvalues have been computed numerically for
different areal densities of flux lines, and as a function of $Q$
(within the irreducible Brillouin zone) and $k_z$.  Several features
emerge from the analysis of our results: {\bf (i)} There is a
weak angular dependence of the harmonic eigenvalues which becomes more
pronounced on approaching the BZ boundary.  {\bf (ii)} Throughout the
BZ, transversal modes are less costly than longitudinal modes, and are
the main cause of lattice fluctuations.  {\bf (iii)} Not surprisingly,
both eigenvalues increase with $\lambda k_z$, and modes with higher
$k_z$ are more costly.  {\bf (iv)} Rather surprisingly, due to a rapid
decrease of $c_{44}^{nl}$, the energy of a transversal mode with a
nonzero $k_z$ actually goes down with $Q$.  The minimum cost occurs
for a finite $Q$ which depends on $\lambda k_z$, and the density of
flux lines.  {\bf (iv)} For a large portion of the IBZ, both
eigenvalues exhibit a linear dependence on $(aQ)^2$, whose slope
depends on the local shear modulus $c_{66}$ which can be calculated as
a function of $a/\lambda$.

Some of the above features are qualitatively well accounted for by the
nonlocal continuum limit results recalculated in
Sec.~\ref{sec:limits}.  Nevertheless, beyond a characteristic value of
$\lambda k_z$, there are major differences between this analytic form
and those calculated numerically.  Guided by our results, we propose
analytic corrections to the nonlocal continuum limit, which fit quite
well the behavior of the elastic eigenvalues throughout the London
regime.  Other limiting forms for the elastic energy often quoted in
the literature, such as the single line, local, and local continuum
approximations, are reconsidered within our framework, and their range
of validity is examined.  We have plotted some of these limits,
together with our numerical results, for better ease of comparison.

We also consider different values of the anisotropy parameter
$\epsilon$, in particular corresponding to values quoted in the
literature for high temperature superconductors such as YBCO or
BiSCCO.  The latter is the paradigm of a highly anisotropic
superconductor with softer elastic eigenvalues.  This fact can be
corroborated in our analysis and is ultimately responsible for
strengthening the magnitude of thermal fluctuations.  We have made use
of our proposed analytical expression to calculate the extent of
thermal fluctuations for different $\epsilon$.  The full form of the
melting curve is then obtained using the Lindeman criterion, capturing
the following salient features of $T_m(B)$: {\bf (i)} It decreases at
high temperatures with the magnetic field, approximately as
$B^{-0.53}$.  {\bf (ii)} It decreases with the anisotropy parameter,
and consequently the molten vortex liquid covers a large fraction of
the equilibrium phase diagram at small $\epsilon$.  {\bf (iii)} There
is reentrant melting at low fields due to the weak interactions
between widely separated flux lines.  The reentrant phase boundary is
itself a non-monotonic function of temperature.  {\bf (iv)} Close to
the critical temperature $T_c$, both branches of the melting line
$B_m$ can be fitted to power laws in the reduced temperature
$1-T/T_c$.  However, the power law is observed at most over two
decades, and the resulting effective exponents are different from the
expectations from mean field theory, casting doubt upon the
effectiveness of this method \cite{Liang96,Welp96,Roulin96}.

Another interesting potential extension of this work is to include
entropic contributions to the free energy, in order to explore the
fluctuation induced effects which have been proposed to be important
in the low field region of the phase diagram \cite{VdW}.  Our analysis
has been applied to a conventional s-wave superconductor with a
triangular lattice of flux lines.  A square lattice of flux lines is
also possible in d-wave superconductors, where similar approach and
phenomenology can be carried out, although the details of the harmonic
energy, as dictated by symmetry, will be different.  Again,
calculation of the entropic contributions to the free energy may
provide a better estimate of the transition between triangular and
square lattices \cite{Maki99}.

\acknowledgements

We are grateful to R. Pastor-Satorras for critical reading of the
manuscript.  This research was supported by grants from the Direcci\'o
General de Recerca (Generalitat de Catalunya), and the National
Science Foundation (Grant No.  DMR-98-05833).

\end{multicols}


\begin{references}
\bibitem{cutoff} Within the London limit, interline distances cannot
be smaller than the coherence length $\xi$ $\sim$ vortex core radius.
That is why a short-distance cutoff of the order of $\xi$ is typically
introduced.
\bibitem{ccutoff} In a continuous description, the coherence length
along the $c$ axis is $\xi_c=\epsilon \xi$. This is also the short-distance
cutoff along the $z$ axis.
\end{references}
\end{document}